\documentclass[journal, 10pt]{IEEEtran}
\IEEEoverridecommandlockouts
\usepackage{cite}
\usepackage{amsmath,amssymb,amsfonts}
\usepackage{algorithmic}
\usepackage{graphicx}
\usepackage{textcomp}
\usepackage{xcolor}
\usepackage{amsfonts}
\usepackage{algorithm}
\usepackage{amsthm}
\usepackage{bm}
\usepackage{subfigure}
\usepackage{pifont}
\usepackage{booktabs, makecell}
\usepackage[utf8]{inputenc}
\usepackage[T1]{fontenc}
\usepackage{upgreek}
\usepackage{mathrsfs}
\usepackage{stfloats}

\DeclareMathOperator{\Var}{Var}

\newcommand{\overbar}[1]{\mkern 1.5mu\overline{\mkern-1.5mu#1\mkern-1.5mu}\mkern 1.5mu}

\def\BibTeX{{\rm B\kern-.05em{\sc i\kern-.025em b}\kern-.08em
    T\kern-.1667em\lower.7ex\hbox{E}\kern-.125emX}}

\usepackage{tikz}
\newcommand*\circled[1]{\tikz[baseline=(char.base)]{
            \node[shape=circle,draw,inner sep=0.8pt] (char) {#1};}}

\newcommand{\bunderbrace}[2]{%
  \begin{array}[t]{@{}c@{}}
  \underbrace{#1}\\
  #2
  \end{array}
}

\newtheorem{lemma}{\textbf{Lemma}}
\newtheorem{theorem}{\textbf{Theorem}}

\newtheorem*{remark}{\textbf{Remark}}

\begin{document}


\title{Power-Domain Interference Graph Estimation for Full-Duplex Millimeter-Wave Backhauling
}

\author{
\IEEEauthorblockN{Haorui Li, Daqian Ding, Yibo Pi, \emph{Member, IEEE}, and Xudong Wang, \emph{Fellow, IEEE}}
\thanks{An earlier version of this paper has been accepted at the IEEE GLOBECOM 2023~\cite{our_paper} and presented in Malaysia in December, 2023.}
\thanks{The authors are with the UM-SJTU Joint Institute, Shanghai Jiao Tong
University, Shanghai 200240, China (email: $\{$haorui.li, daqian.ding, yibo.pi, wxudong$\}$@sjtu.edu.cn).}
}

\maketitle

\begin{abstract}
Traditional wisdom for network resource management allocates separate frequency-time resources for measurement and data transmission tasks. As a result, the two types of tasks have to compete for resources, and a heavy measurement task inevitably reduces available resources for data transmission. This prevents interference graph estimation (IGE), a heavy yet important measurement task, from being widely used in practice. To resolve this issue, we propose to use power as a new dimension for interference measurement in full-duplex millimeter-wave backhaul networks, such that data transmission and measurement can be done simultaneously using the same frequency-time resources. Our core insight is to consider the mmWave network as a linear system, where the received power of a node is a linear combination of the channel gains. By controlling the powers of transmitters, we can find unique solutions for the channel gains of interference links and use them to estimate the interference. To accomplish resource allocation and IGE simultaneously, we jointly optimize resource allocation and IGE with power control. Extensive simulations show that significant links in the interference graph can be accurately estimated with minimal extra power consumption, independent of the time and carrier frequency offsets between nodes.
\end{abstract}


\section{Introduction}
Network densification is a key mechanism to increase spectrum reuse such that the network capacity can be greatly boosted for future explosive growth of user demand. However, denser small cells will inevitably increase the deployment cost if small cells are connected with a fiber backhaul \cite{dense_deployment_cost}. For 
fast and cost-effective deployment of dense small cells, millimeter-wave (mmWave) backhauling is proposed as a promising backhaul alternative. In recent years, mmWave backhauling has been adopted in the integrated access and backhaul (IAB) networks for the deployment of 5G and beyond systems~\cite{iab-survey, TR38874,iab-b5g}, which jointly optimizes the access and backhaul networks for the overall network performance. With the self-interference cancellation (SIC) techniques~\cite{full-duplex-sic,zhang2021design}, full-duplex capability can be integrated for mmWave backhauling. Full-duplex mmWave backhauling can alleviate the network bottleneck, which greatly improves the latency and throughput of IAB networks~\cite{iab-system-level-analysis}.

In mmWave communications, to combat large path loss in high-frequency bands, the transmitters generally employ beamforming techniques to direct their beams towards the receivers, resulting in less interference between links. 
Despite that, under dense deployment, mmWave networks are shown to still operate in the interference-limited regime: the network capacity is significantly affected by the interference among nodes~\cite{interference_limited_regime}. Full-duplex communication allows more concurrent transmissions in the backhaul networks but also incurs more severe and complicated interference~\cite{full-duplex-interference}.
It is thus critical to effectively estimate and manage interference for resource allocation in full-duplex mmWave backhaul networks. However, compared to extensive research on interference management, interference estimation is less explored. Most of the existing works focus on resource allocation under the assumption that interference can be effectively and accurately estimated~\cite{ra_mmwave_iab,  Semi_Centralized_2021}. Our work complements existing efforts on resource allocation by estimating the interference graph that depicts the interference between any pair of mmWave links in the network. Efficient interference graph estimation (IGE) is crucial for the practical use of resource allocation algorithms in full-duplex mmWave backhauling.

In dense mmWave networks, interference is depicted in problem formulations in different forms. The simplest form is a conflict graph (or contention graph) dictating if each pair of links can be active at the same time~\cite{conflict_graph, res_alloc_conflict_graph_2}. Based on the conflict graph, scheduling plans of concurrent transmissions can be determined \cite{resouce_allocation_conflict_graph_wsn, conflict_graph_scheduling_multihop}. In a conflict graph, links not allowed for concurrent transmissions significantly interfere with each other. To construct a conflict graph, we typically need to estimate the interference between each pair of links and compare it against a threshold~\cite{conflict_graph_threshold}. However, as discussed in~\cite{spatial_dl}, it incurs high overheads to directly measure inter-link interference: a network with $n$ links includes $O(n^2)$ pairs of links to be measured. It is possible to reduce the measurement overheads by measuring only a subset of links and inferring the rest using matrix completion~\cite{ige_matrix_completion} or graph embedding techniques~\cite{ige_graph_embed}. However, these techniques leverage node or link similarities to predict interference distribution, incapable of guaranteeing link-level interference estimation accuracy for the construction of conflict graphs.
Moreover, conflict graphs depict the binary relations between links, which cannot capture the additive nature of interference. As a result, conflict graphs cannot effectively avoid significant interference when multiple concurrent links exist.

The second form to characterize interference in mmWave communications is modelling, which can be used to estimate the magnitude of interference and to capture its additive nature. With modelling, the channel and antenna gains for each mmWave link can be easily obtained to estimate the interference. Modelling the channel and antenna patterns requires choosing proper model parameters based on the propagation environment and there exist a variety of choices. A mmWave channel could consist of either a single line of sight (LOS) component, purely scattering components, or a mix of both the LOS and scattering components~\cite{nlos_scatter, mix_los_nlos}. Each component could further experience different degrees of path loss and fading~\cite{TR38901,time_varying_fading}. In outdoor spaces, path loss is typically assumed to be inversely proportional to the square of the distance and frequency, whereas the path loss exponent becomes larger in environment with rich obstacles~\cite{TR38901}. Similarly, the fading channel gain is assumed following different distributions depending on the propagation environment~\cite{time_varying_fading}.
It is, therefore, rather difficult to model a complex propagation environment, which could be best depicted with a combination of dynamic channel models. Moreover, estimating the directional antenna gain often requires approximating the actual antenna patterns, which cannot be simply approximated by a sectored antenna model due to irregular radiation shapes caused by hardware imperfections, as demonstrated in real-world experiments \cite{scaros}. Both the complexities and errors in modelling the channel and antenna patterns make it impractical to estimate the interference accurately. This leads to poor resource allocation decisions and, consequently, degraded network performance.


More recently, considering the complexities in both interference modelling and measurements, machine learning techniques have been used to learn a direct mapping from network information to resource allocation decisions without directly estimating the interference. In~\cite{spatial_dl}, a deep learning approach is proposed to use the geographic locations of nodes as input and learn link scheduling decisions from a large scale of network layouts to achieve the maximum sum rate in device-to-device networks. 
Further improvement using graph embedding and deep neural networks reduces the scale of training data from hundreds of thousands to hundreds of network layouts~\cite{graph_link_schedule}. Similarly, 
deep learning models have also been used to improve other performance metrics, e.g., age of information (AoI)~\cite{aoi_dl}. These machine learning techniques typically learn the mapping from static network information to link scheduling decisions, not adaptive to fast changing network conditions, e.g., fading. 

In cellular networks, to support coordinated multi-point transmission, 4G Long Term Evolution Advanced (LTE-A) standards introduce channel state information interference measurement (CSI-IM)  reference signals to measure the interference between BSs and UEs. Specifically, when an UE reports its received signal strength (RSS) on a resource element (RE) where no signal is transmitted by its serving BS, the reported RSS is then the interference from other BSs on the subcarrier of the RE~\cite{csi_im}. For scheduling purposes, interference has to be measured separately for each interfering BS, which requires operations telling UEs which RBs to measure and which RBs to be used for reporting back measurements. This incurs heavy measurement overheads when the number of interfering BSs is large, especially in dense networks. More importantly, this process requires symbol-level synchronization between BSs and accurate estimation of carrier frequency offsets (CFOs) between UEs and their interfering BSs~\cite{interference_cfo}. However, strict time synchronization is difficult in a fiber backhaul~\cite{time_sync} and becomes even harder in a multi-hop wireless backhaul~\cite{wifiptp}. 

Motivated by these challenges, we want to 1) efficiently estimate the interference graph, depicting the channel gains of interference links, for full-duplex mmWave backhaul networks under time offsets (TO) and CFO, and 2) use the interference graph to improve the efficiency of resource allocation. We propose to use power as a new dimension for interference graph estimation such that data transmission and measurement can be
done simultaneously with the same frequency-time resources. Our core insight is that the received power of a node can be expressed as a linear combination of the channel gains. In a time-slotted network, manipulating the powers of transmitters enables each receiving node to have different received powers across time slots, i.e., a group of linear equations for each node. If the powers of transmitters form a full-rank matrix, the channel gains have a unique solution. Compared to the reference signal (RS)-based IGE, sensitive to TO and CFO, our approach achieves IGE in the power domain, making its performance robust to TO and CFO.

In summary, our contributions in this paper are as follows.
\begin{itemize}
    \item We show that the mmWave network can be considered as a linear system and propose to estimate the interference graph by manipulating the transmit powers of nodes, which is robust to TO and CFO. We further derive the error bounds of the estimated channel gains for both communication and interference links.
    \item We formulate a multi-objective optimization problem to jointly optimize the energy efficiency in resource allocation and the condition number in IGE, such that IGE can be conducted accurately and simultaneously with resource allocation using the same frequency-time resources. We design iterative algorithms to solve the problem and prove the convergence of these algorithms. 
    \item We evaluate our approach with extensive simulations under various network settings and show that our approach can accomplish resource allocation and IGE simultaneously with minimal overhead on power consumption.
\end{itemize}

Note that compared to our previous conference paper~\cite{our_paper}, we have added substantial additional material in this paper, including the practical power control scheme for achieving power linearity, the bounds of channel gain estimation errors, and a new multi-objective formulation and solution of the joint optimization problem for resource allocation and IGE. The rest of this paper is organized as follows. Section \ref{sec:system_model} presents the system model for IGE. Section \ref{section_interference} presents the power-domain approach to IGE and conducts the error analysis. A joint optimization problem is formulated and solved in Section \ref{section:allocation} and the performance is evaluated with simulations in Section~\ref{sec:perf_eval}. Section~\ref{sec:conclusion} concludes the paper.

\section{System Model}
\label{sec:system_model}

This section presents the backhaul network architecture and the channel and interference models. The notations frequently used in this paper are summarized in Table~\ref{table:notation}.

\begin{figure}[!t]
    \centering
        \includegraphics[scale=0.8]{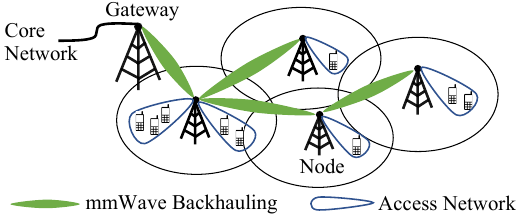}
    \vspace{-0.5em}
    \caption{Multi-hop mmWave backhaul network}\label{fig:network_arch}
    \vspace{-1em}
\end{figure}

\subsection{Millimeter-Wave Backhauling}

We consider a multi-hop mmWave backhaul network with $K$ BSs, one of which serves as a gateway to the core network, as shown in Figure \ref{fig:network_arch}. Each BS is equipped with a full-duplex radio such that it can transmit data to one node and receive data from another node simultaneously. Self-interference cancellation (SIC) techniques are employed at each BS to cancel self-interference.
All BSs are assumed to be time-synchronized and scheduled to transmit and receive data at specified time slots. The scheduling decision is made at the gateway in a centralized way, which requires each BS to report local power measurements of received signals for IGE. After a scheduling plan is made, it is disseminated from the gateway to all BSs. Both the data collection and dissemination for network management purposes are done through the backhaul network using the control channel. Each BS modulates the transmitted symbols with orthogonal frequency division multiplexing (OFDM) and uses all $N_c$ subcarriers for each transmission. Concurrent transmissions in the same time slot may interfere with each other. This time-slotted mmWave backhaul network is compatible with the 5G IAB architecture and thus can be used to provide wireless backhauling in the IAB networks.




We consider the CFO and TO between BSs in the wireless backhaul network. Since BSs are not mobile, the Doppler frequency shift is negligible and CFO is mainly caused by the oscillator mismatch in frequency\cite{interference_cfo}. The TO arises from the synchronization errors among BSs as well as the difference in the propagation delays when considering the signals from multiple transmitters to the same receiver\cite{timing_offset}. We assume that no cooperative transmission is used and that each receiver is assumed to receive data from a single transmitter. For demodulation, each receiver only needs to estimate the CFO of the transmitter and will not compensate the CFOs of other transmitters. Moreover, per the 5G standards~\cite{TS38211}, the subcarrier spacing for mmWave communications could be as large as 960kHz, equating to approximately 1$\mu$s symbol time and 0.07$\mu$s CP length. Since nanosecond-level time synchronization is hard to achieve even in wired networks~\cite{time_sync}, let alone in wireless backhauling networks~\cite{wifiptp}, we assume that TO could be larger than CP length.

\begin{table}[t] 
\centering
\caption{Notations}
\label{Para}
\renewcommand{\arraystretch}{1.35}
\begin{tabular}{|c|l|}
\hline
Notation & Definition \\
\hline
$x_s[i]$ & The $i$-th transmitted sample of node $s$ \\
\hline
$y_s[i]$ & The $i$-th received sample of node $s$ \\
\hline
$X_s[i]$ & The modulated symbol on the $i$-th subcarrier of node $s$ \\
\hline
$X_s^{I,\max}$ & The maximum value of $X_s^I[i]$'s of node $s$ \\
\hline
$X_s^{Q,\max}$ & The maximum value of $X_s^Q[i]$'s of node $s$ \\
\hline
$g_{(s,d),(k,z)}$ & The equivalent channel gain from link $(s,d)$ to $(k,z)$ \\
\hline
$\bm{g}_{(k,z)}$ & Vector of equiv. channel gains from other links to $(k,z)$ \\
\hline
$v[i]$ & The $i$-th sample of noise at a single antenna \\
\hline
$p^{tx}_s[i]$ & The expected transmit power of node $s$ in the $i$-th block\\
\hline
$\bar{p}^{tx}_s[i]$ & The average transmit power of node $s$ in the $i$-th block \\
\hline
$p^{rx}_s[i]$ & The expected received power of node $s$ in the $i$-th block\\
\hline
$\bar{p}^{rx}_s[i]$ & The average received power of node $s$ in the $i$-th block\\
\hline
$N_c$ & Number of subcarriers \\
\hline
$N_g$ & Length of the cyclic prefix\\
\hline
$N_s$ & Number of samples in an OFDM symbol\\
\hline
$\bm{f}_s$ & Errors in the received powers of node $s$ across blocks \\
\hline
\end{tabular}
\label{table:notation}
\end{table}

\subsection{Channel Model}
Let $(s,z)$ be the directional link from BS $s$ to $z$, and $\mathcal{E}$ be the set of all directional links. For simplicity, each BS is assumed to have $M$ transmit antennas and $M$ receive antennas. The channel from BS $s$ to $z$ is denoted as $\mathbf{H}_{(s,z)} \in \mathbb{C}^{M\times M}$. When BS $s$ is transmitting to BS $z$, the transmit beamforming and receive combining vectors of BSs $s$ and $z$ are $\bm{\widetilde{\gamma}}_{s} \in \mathbb{C}^{M\times 1}$ and $\bm{\widetilde{\omega}}_{z} \in \mathbb{C}^{M\times 1}$, respectively.
Since BSs are typically deployed at high places with fixed locations, the channel coherence time between BSs is as large as hundreds of milliseconds for mmWave communications, even in windy days~\cite{coherence_time}. We thus can assume that $\mathbf{H}_{(s,z)}$ is approximately constant during the process of IGE, which takes much less time than the coherence time.\footnote{Our goal is not to estimate $\mathbf{H}_{(s,z)}$ for precoding between the transmitters and receivers, but rather to estimate the gains of interference channels for resource allocation, as detailed in Section \ref{sec:interference_model}.}
We denote the residual CFO of BS $s$ with respect to BS $z$ as $\phi_{(s,z)}$, normalized to the subcarrier spacing with a range between $-0.5$ and $0.5$, and the TO of link $(s,d)$ as $\mu_{(s,d)}$. Let $x_z[i]$ and $y_z[i]$ be the $i$-th transmitted and received samples of nodes $z$, respectively. The $i$-th received sample of node $z$ with beamforming towards node $k$ can be expressed as
\begin{flalign}\label{eq:raw_received_signal}
y_z[i] = &\sum_{(s,d)\in \mathcal{E}\setminus\{(z,k)\}} e^{j2\pi i\phi_{(s,z)}}\bm{\widetilde{\omega}}^T_{z}\mathbf{H}_{(s,z)}\bm{\widetilde{\gamma}}_{s} x_s[i - \mu_{(s,d)}] \notag \\ 
& \quad\quad +  \bunderbrace{\sqrt{\eta}\bm{\widetilde{\omega}}^T_{z}\mathbf{H}_{z}^{SI}\bm{\widetilde{\gamma}}_{z}x_z[i]}{r^{SI}_z[i]} +  \bm{\widetilde{\omega}}^T_{z}\boldsymbol{v}[i],
\end{flalign}
where $\mathcal{E}$ is the set of all links, $(z,k)$ is the directional link originating from node $z$, $\mathbf{H}_{z}^{SI}$ is the self-interference channel, $\eta$ is the SIC capability provided by the SIC techniques, $r^{SI}_z[i]$ is the residual self-interference (SI), and $\boldsymbol{v}[i] \sim \mathcal{CN}(\boldsymbol{0}, \sigma_v^2\boldsymbol{\textnormal{I}})$ is the noise variance at each antenna of the receiving BS. 
Let $h^{\textnormal{eq}}_{(s,d),(k,z)} = e^{j 2\pi i\phi_{(s,z)}}\bm{\widetilde{\omega}}^T_{z}\mathbf{H}_{(s,z)}\bm{\widetilde{\gamma}}_{s}$ be the equivalent channel from link $(s,d)$ to $(k,z)$ 
considering transmit beamforming and receive combining. Similarly, let $h^{SI}_z = \sqrt{\eta}\bm{\widetilde{\omega}}^T_{z}\mathbf{H}_{z}^{SI}\bm{\widetilde{\gamma}}_{z}$ be the equivalent residual SI channel. We can rewrite Eq. (\ref{eq:raw_received_signal}) as
\begin{align}\label{eq:simplified_received_signal}
y_z[i] = \sum_{(s,d) \in \mathcal{E}} h^{\textnormal{eq}}_{(s,d),(k,z)} x_s[i - \mu_{(s,d)}] + n_z[i],
\end{align}
where $h^{\textnormal{eq}}_{(z,d),(k,z)} = h_z^{SI}$ is the equivalent residual SI channel and $n_z[i]=\bm{\widetilde{\omega}}^T_{z}\boldsymbol{v}[i]$. Since beam coherence time is longer than channel coherence time~\cite{beam-coherence-time}, we can consider that $h^{\textnormal{eq}}_{(s,d),(k,z)}$ and $h^{SI}_z$ are approximately constant during the channel coherence time. We want to estimate the magnitudes of these equivalent channels and subsequently use them for resource allocation. 
The time-domain sample $x_s[i]$ is the inverse discrete Fourier transform (IDFT) of the modulated symbols, $\{X_s[r]\}_{r=0}^{N_c-1}$, i.e.,
\begin{equation}
\label{eq:Modulation_with_time_series}
    x_s[i] = \frac{1}{\sqrt{N_c}}\sum_{r=0}^{N_c-1}X_s[r]e^{j2\pi ir/N_c}, 0 \leq i \leq N_c + N_g - 1,
\end{equation}
where $X_s[r]$ is the modulated symbol on subcarrier $r$ transmitted by BS $s$, and $N_g$ is the length of the cyclic prefix (CP).

\subsection{Interference Model}
\label{sec:interference_model}

Resource allocation problems involving interference need to adopt an interference model, where interference is typically assumed to be additive. We present the additive interference model for full-duplex mmWave networks and relate the interference model with the above channel model, such that our IGE approach can be easily integrated with the existing resource allocation problems.
In the mmWave network, each node employs the beamforming techniques to generate a radiation pattern with a main beam and multiple side lobes. Due to hardware constraints, the mmWave nodes could have limited codebook size for beamforming and irregular radiation patterns~\cite{scaros}, which makes interference nulling difficult. We thus consider the interference from both the main beam and the side lobes of transmitters. In mmWave communications, interference depends on the directions of both transmission and reception. Suppose that nodes $s$ and $k$ are transmitting to nodes $d$ and $z$, respectively. Let $g_{(s,d),(k,z)}$ denotes the equivalent channel gain from link $(s,d)$ to link $(k,z)$, where $g_{(z,d),(k,z)}$ is the equivalent residual SI channel gain of node $z$. The equivalent channel $g_{(s,d),(k,z)}$ is a product of the transmission gain at node $s$, the channel gain of the interference link from node $s$ to $z$, and the reception gain at node $z$, denoted as $g_{(s,d),(k,z)}^t$, $g_{(s,d),(k,z)}^c$, and $g_{(s,d),(k,z)}^r$, respectively. 
Let $p^{tx}_s$ be the transmit power of node $s$ to node $d$, and $p^{rx}_z$ be the received power of node $z$ from node $k$. The received power of node $z$ can be expressed as
\begin{equation*}
    p^{rx}_z = \sum_{(s,d) \in \mathcal{E}}\bunderbrace{g_{(s,d),(k,z)}^t g_{(s,d),(k,z)}^c g_{(s,d),(k,z)}^r}{g_{(s,d),(k,z)}} p^{tx}_s + W_z,
\end{equation*}
where $W_z$ is the noise power of node $z$. Suppose that node $k$ is transmitting to node $z$, we can express the signal-to-interference-plus-noise ratio (SINR) at node $z$ as 
\begin{equation}\label{eq:sinr}
    \textnormal{SINR}_{(k,z)} = \frac{g_{(k, z), (k, z)} p^{tx}_k}{W_z + \sum\limits_{(s,d) \in \mathcal{E}\backslash (k,z)} g_{(s, d),(k,z)} p^{tx}_s},
\end{equation}
where $g_{(s,d),(k,z)} = \left|h^{\textnormal{eq}}_{(s,d),(k,z)}\right|^2$.

\section{Interference Graph Estimation: \\ A Power-Domain Approach}
\label{section_interference}

In this section, we present the feasibility of power-domain IGE and design a practical power control scheme. After that, we analyze and bound the channel gain estimation error.

\subsection{Estimating Interference Graph with Power Control}

\begin{lemma}\label{lemma:rx_power_linearity}
In full-duplex mmWave networks, if $\mathbb{E}[|X_k[i]|^2] = \mathbb{E}[|X_k[j]|^2]$ for $i\not=j$ and $\mathbb{E}[X_k[i]] = 0$ for all $i$'s and $k$'s, the expected receive power of node $z$ with beamforming towards node $k$ is a linear combination of the equivalent channel gains, independent of TOs and CFOs, i.e.,
\begin{align}\label{eq:tx_power}
\mathbb{E}[|y_z[i]|^2] = \sum_{(s,d)\in \mathcal{E}}g_{(s,d),(k,z)}\mathbb{E}[|x_s[i]|^2] + W_z,
\end{align}
where $W_z$ is the expected thermal noise power.
\end{lemma}
\begin{IEEEproof}
Please refer to Appendix \ref{section:rx_power_linearity}.
\end{IEEEproof}

Let $s_i$ and $d_i$ be the senders and receiver of the $i$-th directional link, respectively, and $p^{tx}_{s_i}[t]$ and $p^{rx}_{d_i}[t]$ be the transmit and receive powers of nodes $s_i$ and $d_i$ at time $t$.

\begin{theorem}\label{theorem:channel_est}
The equivalent channel gains between links can have a unique solution, if the transmit powers of senders of different links over time satisfy 
\begin{equation}\label{eq:full_rank}
    rank(\mathbf{P}) = |\mathcal{E}|,
\end{equation}
where $\mathbf{P} = \left[\bm{p}^{tx}_{s_1},\dots,\bm{p}^{tx}_{s_{|\mathcal{E}|}}\right]$ and $\bm{p}^{tx}_{s_i} = [p^{tx}_{s_i}[j],\dots, p^{tx}_{s_i}[j+n-1]]^T$ includes the transmit powers of the sender of link $(s_i, d_i)$ from time $j$ to $j+n-1$.
\end{theorem}

\begin{IEEEproof}
Suppose that node $z$ directs beamforming towards node $k$. Let $\bm{p}^{rx}_{z} = \left[p^{rx}_{z}[j], \dots, p^{rx}_{z}[j+n-1]\right]^T$ be the receive powers of node $z$ including the self-interference at different times, and $\textbf{w} = [W_z[j],\dots,W_z[j+n-1]]^T$ be the vector of noise power.
According to Eq. (\ref{eq:tx_power}), we can have
\begin{equation}
    \bm{p}^{rx}_z = \mathbf{P}\boldsymbol{g}_{(k,z)} + \textbf{w},
\end{equation}
where $\boldsymbol{g}_{(k,z)} = [g_{(s_1,d_1), (k,z)}, \dots, g_{(s_{|\mathcal{E}|}, d_{|\mathcal{E}|}), (k, z)}]^T$, $(s_i, d_i) \in \mathcal{E}$, and $g_{(s_i,d_i), (k,z)}$ is the residual SI channel when $s_i = z$.
Since $\mathbf{P} \in \mathbb{R}^{n\times |\mathcal{E}|}$ and $n \geq |\mathcal{E}|$, $\boldsymbol{g}_{(k,z)}$ has a unique solution if $rank(\mathbf{P})= |\mathcal{E}|$. 
\end{IEEEproof}

Theorem \ref{theorem:channel_est} shows that it is feasible to estimate channel gains if the actual transmit powers can approximate the expected transmit powers and form a full-rank matrix. We can also see that compared to the RS-based approach, sensitive to TO and CFO, our approach conducts IGE in the power domain, making its performance robust to TO and CFO. The estimated channel gains can be then used in Eq.~(\ref{eq:sinr}) to estimate the SINR for resource allocation problems.

\subsection{Approximating the Expected Transmit Power}
Suppose that the node $k$ has $N_k$ modulated symbols to send on each subcarrier, equating to $N_k$ OFDM symbols including $N_kN_s$ time-domain samples, where $N_s
= N_c + N_g$ is the number of time-domain samples within an OFDM symbol.
Let $X^I_k[i]$ and $X^R_k[i]$ be the imaginary and real parts of $X_k[i]$, respectively. The maximum values of $X^I_k[i]$ and $X^R_k[i]$ are denoted as $X^{I,\max}_k$ and $X^{Q,\max}_k$. We further denote $x^I_k[i]$ and $x^Q_k[i]$ as the I/Q components of $x_k[i]$ and the average transmit power over $N_k$ OFDM symbols is $\bar{p}^{tx}_k = \frac{1}{N_kN_s}\sum_{j=1}^{N_kN_s}|x_k[j]|^2$ and the expectation of $\bar{p}^{tx}_k$ is denoted as $\mu_k = \mathbb{E}[\bar{p}^{tx}_k]$. We want to approximate $\mu_k$ with $\bar{p}^{tx}_k$ and derive the condition for the difference $\left|\bar{p}^{tx}_k - \mu_k\right|$ to be less than $\mu_k\delta$, where $0 \leq \delta \leq 1$ is the tolerable approximation error.


%

\begin{theorem}
\label{theorem:bound}
Assuming that CP-OFDM is used for modulation and that modulated symbols are chosen with equal probability from the constellation such that $\mathbb{E}[X_k[i]] = 0$ and $\Var(X^I_k[i]) = \Var(X^R_k[i]) = \sigma_k^2$ for all $i$'s and $k$'s, the average transmit power can be bounded as 
\begin{equation*}
    \mathbb{P}\left[\left|\bar{p}^{tx}_k - \mu_k\right| \geq \mu_k\delta \right] \leq 2\min_{\Delta}\left\{e^{f_1(\Delta)} + e^{f_2(\Delta)}\right\},
\end{equation*}
with 
\begin{align*}
    f_1(\Delta) = -\frac{N_kN_g(N_c^2-N_c)\sigma_k^4}{4(X^{I,\max}_k)^4}Q\left(\frac{(X^{I,\max}_k)^2N_s\Delta}{N_g(N_c-1)\sigma_k^4}\right),
\end{align*}
and 
\begin{equation*}
    f_2(\Delta) = -\frac{N_kN_c\eta_k^2}{b_k^2}Q\left(\frac{b_k(\mu_k\delta - \Delta)}{\eta_k^2}\right),
\end{equation*}
where $\Delta$ is chosen to minimize the sum of $e^{f_1(\Delta)}$ and $e^{f_2(\Delta)}$, $Q(t) = (1+t)\log(1+t)$, $b_k = 2\left(X^{I,\max}_k\right)^2 - 2\sigma_k^2$, and $\eta_k^2 = \mathbb{E}\left[|X_k[i]|^4\right] - 4\sigma_k^4$. 
\end{theorem}

\begin{IEEEproof}
    Please refer to the Appendix \ref{section:proof_theorem_bound}.
\end{IEEEproof}

\begin{remark}
For modulations with symbols of equal energy, e.g., quadrature phase shift keying (QPSK) and binary phase-shift keying (BPSK), the average power $\bar{p}^{tx}_k$ is equal to $\mu_k$, indicating that $\mathbb{P}\left[\left|\bar{p}^{tx}_k - \mu_k\right| \geq \mu_k\delta \right] = 0$ for $\delta > 0$.
\end{remark}

\begin{figure}[t]
    \centering
    \subfigure[$\delta = 0.01$]{
        \label{fig:delta=0.01}
        \includegraphics[scale=0.35]{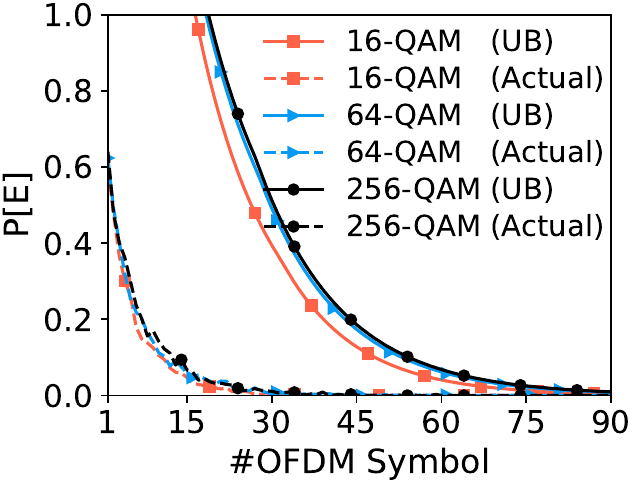}
    }
    \subfigure[$\delta = 0.02$]{
        \label{fig:delta=0.02}
        \includegraphics[scale=0.35]{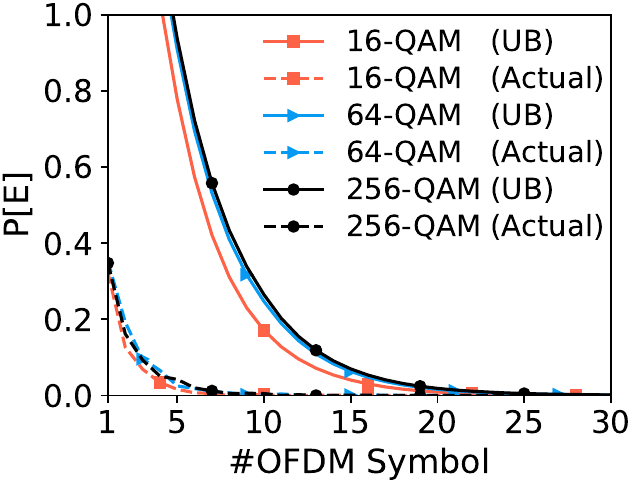}
    }
    \caption{Error bounds under square $M$-QAM modulations, where $\mathbb{P}[E]$ denotes the probability of the approximation error falling beyond the tolerable range.}
    \label{fig:error_bound}
    \vspace{-1em}
\end{figure}

Using Theorem \ref{theorem:bound}, we show that it is feasible to approximate the expected transmit power with a small number of OFDM symbols. Let $E$ denote the event that $\left|\bar{p}^{tx}_k - \mu_k\right|$ is greater than $\mu_k\delta$. We want $\mathbb{P}[E]$ to be controlled within a tolerable range. Fig. \ref{fig:error_bound} shows the simulation results for $\mathbb{P}[E]$'s under different square $M$-quadrature amplitude modulation ($M$-QAM) modulations, where $N_c = 1024$ and $N_g = 72$. When $\delta = 0.01$, the solid lines in Fig. \ref{fig:delta=0.01} show that more than 78 OFDM symbols are needed to reduce the upper bound (UB) of $\mathbb{P}[E]$ to 0.01 or lower for 16-QAM modulation. For 64-QAM modulation, this number increases to 86 OFDM symbols. From 64-QAM to 256-QAM modulations, the UBs are almost the same. To understand $\mathbb{P}[E]$ in random cases, we randomly select modulated symbols with equal probability from the constellation and check if event $E$ occurs. This process is repeated 10,000 times under each symbol size (i.e., the number of OFDM symbols used) to calculate $\mathbb{P}[E]$. The dashed lines show that the probability of event $E$ in random cases is much less than the upper bound. Using only about 25 OFDM symbols can control $\mathbb{P}[E]$ to 0.01 or lower. Relaxing the approximation error ($\delta$) to $0.02$ can further reduce the required number of OFDM symbols to 5 for achieving the same $\mathbb{P}[E]$. From the above results, we can conclude that it is possible to approximate the expected transmit power with a reasonable number of OFDM symbols.

\begin{figure}[!t]
    \centering
        \includegraphics[scale=0.45]{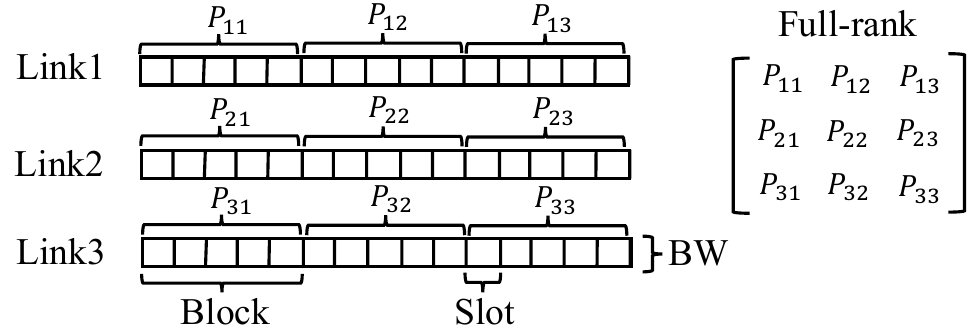}
    \caption{Power control scheme for interference graph estimation}\label{fig:power_control_scheme}
    \vspace{-1em}
\end{figure}

\subsection{Power Control Scheme for Interference Graph Estimation}

Theorems \ref{theorem:channel_est} and \ref{theorem:bound} together indicate that we can design a simple power control scheme for IGE in a time-slotted mmWave multi-hop network. As shown in Fig. \ref{fig:power_control_scheme}, to approximate the expected transmit power, each link has to transmit with the same power over several consecutive time slots, referred to as a \emph{block}. The number of consecutive time slots (denoted as $n_{s}$) in a block is determined based on Theorem \ref{theorem:bound}. Since links may differ in modulation schemes, blocks of different lengths are needed for each link. For design simplicity, we enforce all links to use the same block length. Since this example includes three links, a 3-by-3 full-rank transmit power matrix needs to be constructed. Suppose that a network includes $n_{l}$ links. A total number of $n_{s}n_{l}$ slots is required for the IGE process. Reducing $n_{s}$ helps accelerate the IGE process at the cost of an increasing approximation error to the expected transmit power. The trade-off between the time consumption and accuracy of IGE will be further discussed in Section \ref{sec:design_param}.

From Theorem \ref{theorem:channel_est}, we know that the expected received power is a linear combination of the equivalent channel gains and the expected transmit powers, i.e., $\bm{p}^{rx}_z=\mathbf{P}\bm{g}_{(k,z)}+\textbf{w}$, which can be rewritten as
\begin{equation}\label{eq:perfect_linearity}
  \mathbf{P}\bm{g}_{(k,z)}=\bm{b}_z,
\end{equation}
where $\bm{b}_z=\bm{p}^{rx}_z-\textbf{w}$. However, as the number of OFDM symbols used for transmission is limited, $\bm{b}_z$ is not directly measurable. The receiver can only measure the actual received power, denoted as $\bm{\overbar{b}}_z$, by averaging over the received symbols in each block.  
We estimate the channel gains as
\begin{equation}\label{eq:perturbation linearly}
\mathbf{P}\bm{\hat{g}}_{(k,z)}=\bm{\overbar{b}}_z,
\end{equation}
where $\bm{\hat{g}}_{(k,z)}= [\hat{g}_{(s_1,d_1), (k,z)}, \dots, \hat{g}_{(s_{|\mathcal{E}|}, d_{|\mathcal{E}|}), (k, z)}]^T$ denotes the estimated equivalent channel gains from other links to link $(k,z)$. 

\subsection{Error Analysis}
\label{sec:error_analysis}


This section presents two types of channel estimation errors and their bounds. The \emph{per-link error} is the difference between $\bm{g}_{(k,z)}$ and $\bm{\hat{g}}_{(k,z)}$, which describes the channel estimation errors between link $(k,z)$ and each of the rest. The \emph{per-pair error} is the channel estimation error between two specific links, which is of more granularity. Both the estimation errors will be considered in the formulation of the resource allocation problems in Section~\ref{section:allocation}.

Suppose that $n_b$ blocks are used for the IGE process, where $n_b \geq |\mathcal{E}|$. Let $\bm{f}_z \in \mathbb{R}^{n_b \times 1}$ be the difference between the actual and the expected received powers of node $z$, i.e., $\bm{\overbar{b}}_z = \bm{b}_z + \bm{f}_z$.  Let $\epsilon_{\bm{b}} =||\bm{f}_z||\slash||\bm{b}_z||$ be the relative perturbation for the received power. 




\begin{theorem}
\label{theorem:cond_number_bound}
(Per-link error) The channel estimation error for link $(k,z)$ can be bound as
\begin{align*}
    \lVert\hat{\bm{g}}_{(k,z)}-\bm{g}_{(k,z)}\rVert \leq \kappa(\mathbf{P})\lVert\bm{g}_{(k,z)}\rVert\epsilon_{\bm{b}}
\end{align*}
where $\kappa(\mathbf{P})$ is the condition number of $\mathbf{P}$. 
\end{theorem}
    
\begin{IEEEproof}
Recall that $\mathbf{P} \in \mathbb{R}^{n_b\times |\mathcal{E}|}$ and $n_b \geq |\mathcal{E}|$, where $\mathbf{P}$ is a column full-rank matrix with the left inverse $\mathbf{P}^{\dag}$. Combining Eq. (\ref{eq:perfect_linearity}) and (\ref{eq:perturbation linearly}), we have $\hat{\bm{g}}_{(k,z)}-\bm{g}_{(k,z)}=\mathbf{P}^{\dag}\bm{f}_{(k,z)}$. Hence,
\begin{align*}
\frac{\lVert\hat{\bm{g}}_{(k,z)}-\bm{g}_{(k,z)}\lVert}{\lVert\bm{g}_{(k,z)}\lVert} &\leq \frac{\lVert\mathbf{P}^{\dag}\lVert\lVert\bm{f}_z\lVert}{\lVert\bm{g}_{(k,z)}\lVert} \leq\lVert\mathbf{P}^{\dag}\lVert\lVert\mathbf{P}\lVert \frac{\lVert\bm{f}_z\lVert}{\lVert\mathbf{P}\lVert\lVert\bm{g}_{(k,z)}\lVert}\\
&\leq\lVert\mathbf{P}^{\dag}\lVert\lVert\mathbf{P}\lVert \frac{\lVert\bm{f}_z\lVert}{\lVert\bm{b}_z\lVert} = \kappa(\mathbf{P})\epsilon _{\bm{b}}.
\end{align*}
\end{IEEEproof}

Theorem \ref{theorem:cond_number_bound} indicates that we can lower the channel estimation bound for link $(k,z)$ by reducing either $\kappa(\mathbf{P})$ or $\epsilon_{\bm{b}}$. We can minimize $\kappa(\mathbf{P})$ by carefully choosing the transmit powers of nodes, as will be further discussed in Section~\ref{section:allocation}. We can further decrease $\kappa(\mathbf{P})$ by using a transmit power matrix of a larger size.

\begin{theorem}
\label{theo:cn_block}
If the transmit powers of nodes can be controlled, for $\mathbf{P} \in \mathbb{R}^{n_b\times |\mathcal{E}|}$, the condition number of $\mathbf{P}$ can be reduced by using a larger $n_b$.
\end{theorem}

\begin{IEEEproof}
Let $\mathbf{P}_{i} = [\bm{p}^{tx}[1],\dots,\bm{p}^{tx}[j],\dots,\bm{p}^{tx}[i]]^T$ be a submatrix consisting of the first $i$ rows of $\mathbf{P}$, where $\bm{p}^{tx}[j]=[p^{tx}_{s_1}[j],\dots,p^{tx}_{s_{|\mathcal{E}|}}[j]]$ and $i = |\mathcal{E}|,\dots,n_b$. Let $\alpha_i$ and $\beta_i$ be the largest and smallest singular values of $\mathbf{P}_{i}$, respectively, and $k_{i} = \alpha_i/\beta_i$ be the condition number. It has been shown in \cite{dax2023kaczmarz} that $\alpha_{i+1}^2 \leq \alpha_i^2 + (\bm{u}_{i+1}^T\bm{p}[i+1])^2$ and $\beta_{i+1}^2 \geq \beta_i^2 + (\bm{v}_{i+1}^T\bm{p}[i+1])^2$,
where $\bm{u}_{i+1}$ and $\bm{v}_{i+1}$ are the eigenvectors of $\mathbf{P}_{i+1}^T\mathbf{P}_{i+1}$ corresponding to $\alpha_i^2$ and $\beta_i^2$, respectively. Hence,
\begin{equation*}
k_{i+1}^2 \leq k_i^2 \times \frac{1 + (\bm{u}^T_{i+1}\bm{p}^{tx}[i+1])^2/\alpha_i^2}{1 + (\bm{v}^T_{i+1}\bm{p}^{tx}[i+1])^2/\beta_i^2}.
\end{equation*}
Since $\alpha_i^2 \geq \beta_i^2$, if $\bm{v}^T_{i+1}\bm{p}^{tx}[i+1] > \bm{u}^T_{i+1}\bm{p}^{tx}[i+1]$, we can conclude $k_{i+1} < k_i$. The condition can be satisfied if we let the transmit power matrix $\mathbf{P}_{i+1} = [\mathbf{P}_i; \bm{p}^{tx}[i+1]]$, where $\bm{p}^{tx}[i+1] = \bm{v}_{i+1}$. This means that using a larger transmit matrix helps reduce the condition number of $\mathbf{P}$.
\end{IEEEproof}

\begin{table*}[t]
\centering
\begin{equation}\label{eq:f}
 f_z[i]=\underbrace{\frac{1}{m}\sum_{r=(m-1)i+1}^{mi} \left(F_{(k,z)}[r] + V_{(k,z)}[r]\right)}_{\circled{\small 1}} + \underbrace{\frac{1}{m}\sum_{r=(m-1)i+1}^{mi}s[r]}_{\circled{\small 2}} 
+ \underbrace{\frac{1}{m}\sum_{r=(m-1)i+1}^{mi}|n_z[r]|^2}_{\circled{\small 3}}, i \in \{1,\dots, n_b\}
\end{equation}
\hrule
\end{table*}

We will show how the choice of $n_b$ decreases the condition number of $\mathbf{P}$ with experiments in Section~\ref{sec:design_param}. 

\begin{lemma}\label{lemma:power_perturbation_one_channel}
Assuming that $\bm{f}_z$ is known, the estimation error for the equivalent channel gain between a pair of links can be bounded as
\begin{equation}
|\hat{g}_{(s, d), (k,z)}-g_{(s, d), (k,z)}| \leq \lVert\bm{r}_{i}\lVert\cdot\lVert\bm{f}_z\rVert,
\end{equation}
where $i$ is the index of $g_{(s, d), (k,z)}$ in $\bm{g}_{(k,z)}$ and $\bm{r}_{i}$ is the $i$-th row of $\mathbf{P}^{\dag}$, the left inverse of $\mathbf{P}$.
\end{lemma}

\begin{IEEEproof}
Recall that we want to estimate $\bm{\hat{g}}_{(k,z)}$ from the linear system,
$\mathbf{P}\bm{\hat{g}}_{(k,z)} = \bm{b}_z + \bm{f}_z$. Based on the perturbation theory in \cite{chandrasekaran1991perturbation}, the component-wise error can be written as 
$\left|\hat{g}_{(s, d), (k,z)}-g_{(s, d), (k,z)}\right|=\lVert\bm{r}_{i}\rVert\lVert\bm{b}\rVert\epsilon_{\bm{b}}\cos\boldsymbol{\psi_{\bm{f},i}}$,
where $\cos\boldsymbol{\psi_{f,i}}$ is the angle between $\bm{r}_{i}$ and $\bm{f}_z$. Hence,
\begin{align*}
\left|\hat{g}_{(s, d), (k,z)}-g_{(s, d), (k,z)}\right| &=\lVert\bm{r}_{i}\rVert\lVert\bm{f}_z\rVert \left|\cos\boldsymbol{\psi_{\bm{f},i}}\right| \\
&\leq \lVert\bm{r}_{i}\rVert\lVert\bm{f}_z\rVert.
\end{align*}
\end{IEEEproof}

Lemma \ref{lemma:power_perturbation_one_channel} tells us that the upper bound of the component-wise error depends on $\lVert\bm{f}_z\rVert$. By further bounding $\lVert\bm{f}_z\rVert$, we can derive a bound for the estimation error. However, bounding $\lVert\bm{f}_z\rVert$ requires knowing $\bm{g}_{(k,z)}$, which cannot be obtained. Instead, we want to bound the expected estimation error for the equivalent channel gain, denoted as $\Delta_{(s,d),(k,z)} = \mathbb{E}_g[|\hat{g}_{(s, d), (k,z)}-g_{(s, d), (k,z)}|]$.

To bound $\Delta_{(s,d),(k,z)}$, we write $f_z[i]$ as in Eq. (\ref{eq:f}),
where 
\begin{align}
&F_{(k,z)}[r] = \sum_{(s,d)\in\mathcal{E}}\sum_{\substack{(l,v)\in\mathcal{E} \notag \\ (l,v)\not=(s,d)}} q_{(s,d), (k,z)}[r]q^*_{(l,v),(k,z)}[r], \notag \\
&V_{(k,z)}[r]=n_z[r] \sum_{(s,d)\in\mathcal{E}} q_{(s, d),(k,z)}[r], \notag \\
&s[r]=\sum_{(s,d)\in\mathcal{E}} g_{(s,d),(k,z)} 
\left(\mathbb{E}[|x_s[r]|^2]- |x_s[r]|^2\right), \notag \\
&q_{(s, d),(k,z)}[r] = h^{\textnormal{eq}}_{(s,d),(k,z)} x_{(s,d)}[r - \mu_{(s,d)}]. \notag 
\end{align}
We want to bound the three items in Eq. (\ref{eq:f}) separately. Suppose that $m_i$ samples are needed for the probability of items $\circled{\small $i$}$ being greater than $A_i$ to be less than $\beta_i$, i.e., $\mathbb{P}[|\circled{\small $i$}| > A_i] \leq \beta_i$, for $i = 1, 2, 3$. We can calculate $A_i$'s as follows.

\begin{lemma}
\label{eq:f_upper_bound}
For edge $(k, z) \in \mathcal{E}$, given $\beta$ and the number of samples, $m_i$'s, $A_i$'s can be calculated as
\begin{align*}
&A_1^2 = \frac{4}{m_1\beta_1}\sum_{(s,d)\in\mathcal{E}}\sum_{\substack{(l,v)\in\mathcal{E} \\ (l,v)\not=(s,d)}}\sigma^2_s\sigma^2_l\mathbb{E}\left[g_{(s,d),(k, z)}\right]\mathbb{E}\left[g_{(l,v),(k, z)}\right], \notag \\
&A_2^2 = \frac{1}{m_2\beta_2}\sum_{(s,d)\in\mathcal{E}}\mathbb{E}\left[g_{(s,d),(k,z)}^2\right](4\sigma_s^4-\frac{4}{N_c}\sigma_{s}^4), \notag \\
&A_3^2 = \frac{1}{m_3\beta_3}\sigma_v^4. \notag
\end{align*}
\end{lemma}

\begin{IEEEproof}
Please refer to Appendix \ref{appendix:f_upper_bound}.
\end{IEEEproof}

With the probabilistic bound for each item in Eq. (\ref{eq:f}), the \emph{per-pair estimation error} can then be bounded as follows. The per-pair error will be used in the optimization problem ($\mathcal{P}1$) to incorporate the impact of channel gain estimation error.

\begin{theorem}
(Per-pair error) For edge $(k, z) \in \mathcal{E}$, the expected channel gain estimation error, $\Delta_{(s,d),(k,z)}$, can be bounded with probabilistic guarantee as
\begin{equation*}
        \mathbb{P}\left[\Delta_{(s,d),(k,z)} \leq \lVert \bm{r}_i \rVert\sum_{j=1}^3 A_j\right] \geq 1 - \beta,
\end{equation*}
where the number of samples used is the maximum of $m_1$, $m_2$, and $m_3$, and $\beta = 1 - \prod_{j=1}^3 (1 - \beta_j)$.
\end{theorem}

\begin{IEEEproof}
We know that $\mathbb{P}[|\circled{\small \(j\)}| \leq A_j] \geq 1 - \beta_j$. Hence,
\begin{align}
\mathbb{P}\left[\Delta_{(s,d),(k,z)} \leq \lVert\bm{r}_i\rVert \sum_{j=1}^3 A_j \right] &\geq \prod_{j=1}^3\mathbb{P}\left[|\circled{\small \(j\)}| \leq A_j \right] \notag \\
&\geq \prod_{j=1}^3 (1 - \beta_j)  \notag
\end{align}
Given the probabilistic guarantee, $1 - \beta$, we can calculate the $\beta_j$'s that must be satisfied by each item.
\end{IEEEproof}

\section{Joint Optimization for Interference Graph Estimation and Resource Allocation}
\label{section:allocation}

As our IGE approach is in the power domain, we can jointly optimize the accuracy of IGE and the network performance with power control. We consider the impact of estimation errors on resource allocation in the problem formulation, such that our power allocation scheme is robust to estimation errors.

\subsection{Problem Formulation}
Since each node is equipped with a single full-duplex radio, each node has a single path to the gateway, determined by the routing algorithm. Based on the buffer status reports of UEs, we can infer the traffic demand for each link. Our goal is to 1) maximize the energy efficiency of the mmWave backhaul network in satisfying the traffic demand within the required time and 2) minimize estimation errors. As indicated by Theorem \ref{theorem:cond_number_bound}, the per-link error can be reduced by decreasing the condition number of the transmit power matrix. Let $\mathcal{I}$ be the set of indices for blocks and $\mathcal{N}$ be the set of indices for nodes. We can formulate a multi-objective joint optimization problem as follows.
\begin{align}
(\mathcal{P}1) \ \min_{\mathbf{P}, \bm{\delta}} \quad & \left[\sum_{i\in\mathcal{I}}\sum_{(k,z) \in \mathcal{E}} p^{tx}_k[i]\tau, \kappa(\mathbf{P}) \right] \notag \\
\textrm{s.t.} \quad & (C_1)\ \sum\nolimits_{k \in \mathcal{N}} \delta_{(k, z)}[i] \leq 1, \forall i \in \mathcal{I}, \forall z \in \mathcal{N} \notag \\
& (C_2)\ \sum\nolimits_{z \in \mathcal{N}} \delta_{(k, z)}[i] \leq 1, \forall i \in \mathcal{I}, \forall k \in \mathcal{N}  \notag \\
& (C_3)\ \delta_{(k,z)}[i] \in \{0, 1\}, \forall (k,z) \in \mathcal{E}, \forall i \in \mathcal{I} \notag \\
& (C_4)\ \sum_{i = 1}^{|\mathcal{I}|} \delta_{(k, z)}[i] \geq \left\lceil \frac{D_{(k, z)}}{R_{(k, z)}\tau} \right\rceil, \forall (k, z) \in \mathcal{E}, \notag \\
& (C_5)\ \widehat{\text{SINR}}_{(k,z)} \geq \gamma_{(k,z)} \delta_{(k,z)}[i], \forall (k, z) \in \mathcal{E},\notag \\
& (C_6)\ p^{tx}_k[i] \geq \delta_{(k,z)}[i]P_{min}, \forall (k, z) \in \mathcal{E}, \notag 
\\
& (C_7)\ p^{tx}_k[i] \leq \delta_{(k,z)}[i]P_{max}, \forall (k, z) \in \mathcal{E}, \notag \\
&  (C_8)\ rank(\mathbf{P}) = |\mathcal{E}|, \notag
\end{align}
where $\tau$ is the block length, $\delta_{(k, z)}[i]$ indicates if node $k$ is sending to node $z$ at the $i$-th block, $D_{(k,z)}$ is the traffic demand for link $(k,z)$, $R_{(k,z)}$ is the transmission rate under the selected modulation and coding scheme (MCS), $\gamma_{(k, z)}$ is the required SINR by link $(k, z)$ to support the transmission rate $R_{(k, z)}$, $P_{max}$ and $P_{min}$ are the boundary powers, and $\widehat{\text{SINR}}_{(k,z)}$ is the SINR calculated using the conservative channel gains that consider the estimation errors for link $(k,z)$. We define $\widehat{\text{SINR}}_{(k,z)}$ as
\begin{equation}\label{eq:sinr_estimated}
    \widehat{\textnormal{SINR}}_{(k,z)} = \frac{g_{(k, z), (k, z)} p^{tx}_k}{W_z + \sum\limits_{(s,d) \in \mathcal{E}\backslash (k,z)} g_{(s, d),(k,z)}^{ub} p^{tx}_s},
\end{equation}
where $g_{(k,z),(k,z)}$ is the equivalent channel gain for the communication link from node $k$ to $z$, $g_{(s, d),(k,z)}^{ub} = \hat{g}_{(s, d),(k,z)} + \Delta_{(s, d),(k,z)}$, and $\hat{g}_{(s, d),(k,z)}$ is the estimated equivalent channel gain for the interference channel from link $(s,d)$ to $(k,z)$. For each communication link, the channel gain will be estimated for the demodulation of received symbols. We simply assume that $g_{(k, z),(k,z)}$ be accurately estimated for the communication links, which can be easily extended to account for the estimation errors. The equivalent channel gain, $\hat{g}_{(s, d),(k,z)}$, is estimated based on Eq. (\ref{eq:perturbation linearly}) with the received powers measured in the previous round of scheduling.

\subsection{Proposed Solution}

\begin{figure}[!t]
    \centering
        \includegraphics[scale=0.45]{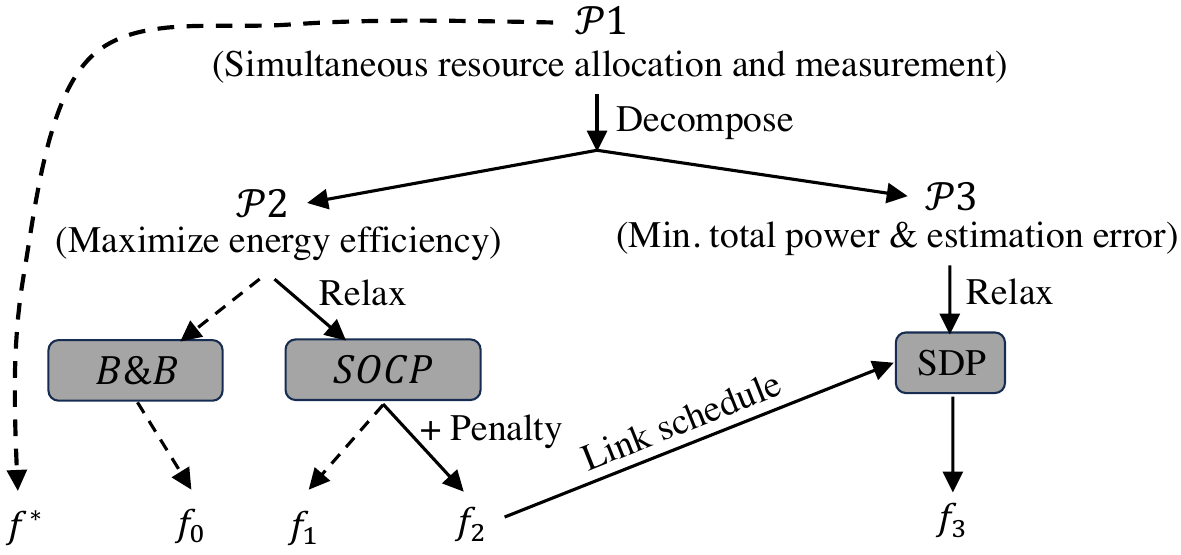}
    \caption{Workflow of our proposed solution}\label{fig:workflow}
    \vspace{-1em}
\end{figure}

$\mathcal{P}1$ is a multi-objective mixed integer nonlinear programming (MINLP) problem with a nonlinear objective function and a rank constraint. When the decision space is large (with hundreds of binary and continuous variables), it is challenging to find the optimal solution for $\mathcal{P}1$ due to the high computational complexity. We thus propose a heuristic approach with the workflow shown in 
Fig. \ref{fig:workflow}. Specifically, we decompose $\mathcal{P}1$ into two subproblems $\mathcal{P}2$ and $\mathcal{P}3$.
$\mathcal{P}2$ is an MINLP problem aiming to maximize the energy efficiency while satisfying user traffic demand, i.e.,
\begin{align}
(\mathcal{P}2) \ \min\limits_{\bm{P}, \boldsymbol{\delta}} &\sum_{i\in\mathcal{I}}\sum_{(k,z) \in \mathcal{E}} p_k^{tx}[i] \nonumber \\
\text{s.t.} \ & C_1\mbox{-}C_7, \nonumber
\end{align}
where the link schedule $\boldsymbol{\delta}$ including the on/off states of links in each block is determined. $\mathcal{P}3$ takes as input the link schedule $\boldsymbol{\delta}$ from $\mathcal{P}2$ to minimize two objectives, the total power consumption and estimation error. $\mathcal{P}3$ can be written as a linear scalarization problem to combine the two objectives using weights, i.e.,
\begin{align}
(\mathcal{P}3) \ \min\limits_{\bm{P}} & \ \alpha_1\frac{1}{\beta_1}\kappa(\mathbf{P})+\alpha_2\frac{1}{\beta_2
}\sum_{i\in\mathcal{I}} \sum_{(k,z) \in \mathcal{E}} p^{tx}_k[i] \nonumber \\
\text{s.t.} \ & C_5 \mbox{-} C_8, \nonumber
\end{align}
where variable $\alpha_i > 0$, $\sum_i \alpha_i = 1$, specifies the priority of the $i$-th objective, and $\beta_1$ and $\beta_2$ are the normalization factors for the condition number and power consumption. Varying $\alpha_i$'s allows us to obtain a set of Pareto optimal solutions on the Pareto front. 
Based on $\boldsymbol{\delta}$, $\mathcal{P}3$ obtains the power allocation $\mathbf{P}$ that satisfies the rank constraint and minimize the weighted condition number. $\mathcal{P}3$ differs from $\mathcal{P}2$ in that as $\boldsymbol{\delta}$ is given in $\mathcal{P}3$, its variable list does not include $\boldsymbol{\delta}$, making it a problem with only continuous variables.

$\mathcal{P}2$ will be relaxed into a second-order cone programming (SOCP) problem with items penalizing the binary decision variables deviating from zero or one, and $\mathcal{P}3$ will be relaxed into a semi-definite programming (SDP) problem, where $f_2$ is the objective value for solving the relaxed $\mathcal{P}2$ with penalty items and $f_3$ is the objective value of our heuristic solution for $\mathcal{P}1$. For optimality loss, we want to evaluate the performance gap between the optimal and the heuristic solutions. In Fig. \ref{fig:workflow}, $f^*$ is the optimal objective value for $\mathcal{P}1$, $f_0$ is the optimal objective value for $\mathcal{P}2$, and $f_1$ is the objective value for relaxed $\mathcal{P}2$ with no penalty terms. We know that these values can be ordered as $f_1 \leq f_0 \leq f^* \leq f_3$.\footnote{Since $f_0$ are $f_1$ are the objective values for the original and relaxed $\mathcal{P}2$, respectively, we have $f_1 \leq f_0$. Since $\mathcal{P}1$ has one more constraint (i.e., the rank constraint) than ${P}2$, we expect that $f_0 \leq f^*$. Moreover, as $f_3$ is the objective value of the heuristic solution for $\mathcal{P}1$, we know that $f^* \leq f_3$.} For small-sized problems, $f_0$ can be obtained with the branch-and-bound (B\&B) approach. In Section \ref{sec:perf_eval}, we will use the performance gap between $f_0$ and $f_3$ to evaluate the performance gap between the optimal and our heuristic solution.

\subsection{Solution to $\mathcal{P}2$}





We want to approximately solve $\mathcal{P}2$ with continuous and convex relaxations. We first relax the binary variables $\delta_{(k,z)}[i]$'s in $\mathcal{P}2$ into continuous variables between 0 and 1. As a result, the problem becomes a nonconvex nonlinear problem. After the problem is solved, we want to convert relaxed continuous $\delta_{(k,z)}[i]$'s back to binary and thus introduce a penalty term in the objective function to penalize $\delta_{(k,z)}[i]$'s for deviating from 0 or 1. The penalty term is
$f(\boldsymbol{\delta}) = g_1(\boldsymbol{\delta}) + g_2(\boldsymbol{\delta})$, where $g_1(\boldsymbol{\delta}) = \sum\limits_{i\in\mathcal{I}}\sum_{(k,z) \in \mathcal{E}} \delta_{(k,z)}[i]$ and $g_2(\boldsymbol{\delta}) = -\sum\limits_{i\in\mathcal{I}}\sum_{(k,z) \in \mathcal{E}} (\delta_{(k,z)}[i])^2$.
Suppose $\boldsymbol{\delta}^{(t-1)}$ is a feasible solution at the $(t-1)$-th iteration. $g_2(\boldsymbol{\delta})$ can be linearized by its first-order Taylor approximation near $\boldsymbol{\delta}^{(t-1)}$ as 
\begin{equation*}
g_2(\boldsymbol{\delta}) \leq \Tilde{g_2}(\boldsymbol{\delta}) \triangleq g_2(\boldsymbol{\delta}^{(t-1)}) + \nabla g_2^T(\boldsymbol{\delta}^{(t-1)})(\boldsymbol{\delta} - \boldsymbol{\delta}^{(t-1)}).
\end{equation*}


Next, we want to relax the nonconvex constraint $C_5$ into a convex one and use the successive convex approximation (SCA) technique to iteratively solve the convex problem until convergence.
To relax the constraint $C_5$, we divide it into two constraints, i.e.,
\begin{equation}
(C_{5.1}) \ a_{(k,z)}[i] = BN_0 + \sum_{(l,d) \in \mathcal{E}\backslash (k,z)} g_{(l, d),(k,z)} p^{tx}_l[i], 
\end{equation}
and
\begin{equation}
(C_{5.2}) \ \frac{p_{(k,z)}^{tx}[i]}{\gamma_{(k,z)}} g_{(k,z),(k,z)} \geq \delta_{(k,z)}[i]a_{(k,z)}[i].
\end{equation}
Following the principle of SCA, we need to find a convex surrogate function for the nonconvex term in $C_{5.2}$ as
\begin{equation}\label{eq:surrogate_equality}
   \delta_{(k,z)}[i]a_{(k,z)}[i] \leq  \frac{\phi_{(k,z)}[i]}{2}(\delta_{(k,z)}[i])^2 + \frac{(a_{(k,z)}[i])^2}{2\phi_{(k,z)}[i]}
\end{equation}
for any constant $\phi_{(k,z)}[i] > 0$, where the equality holds when $\phi_{(k,z)}[i] = a_{(k,z)}[i] / \delta_{(k,z)}[i]$. Then, constraint $C_{5.2}$ becomes
\begin{align*}
(C_{5.3})\ \frac{g_{(k,z),(k,z)}}{\gamma_{(k,z)}}p_{(k,z)}^{tx}[i] \geq \frac{\phi_{(k,z)}[i]}{2}(\delta_{(k,z)}[i])^2 + \frac{(a_{(k,z)}[i])^2}{2\phi_{(k,z)}[i]},
\end{align*}
which is a convex constraint.

Let $\boldsymbol{A}$ be the matrix for $a_{(k,z)}[i]$'s. At the $t$-th iteration, $\mathcal{P}2$ can then be relaxed as
\begin{align}
(\mathcal{P}2^*) \ \min\limits_{\boldsymbol{P}, \boldsymbol{\delta}, \boldsymbol{A}} &\sum_{i \in \mathcal{I}}\sum\limits_{(k,z) \in \mathcal{E}} p_k^{tx}[i] + \lambda g_1(\boldsymbol{\delta}) + \lambda \Tilde{g}_2(\boldsymbol{\delta})  \nonumber \\
\text{s.t.} \ & C_1, C_2, C_4, C_{5.1},C_{5.3}, C_6, C_7, \notag \\
              & \delta_{(k,z)}[i] \in [0, 1], \forall (k,z)\in \mathcal{E}, \forall i \in \mathcal{I},
\nonumber
\end{align}
which is a SOCP problem. The solution to $\mathcal{P}2$ can be obtained by iteratively solving $\mathcal{P}2^*$ until convergence.

\subsection{Solution to $\mathcal{P}3$}
When $\boldsymbol{\delta}$ is given, constraint $C_5$ becomes linear. 
As shown in \cite{b22}, rank constraints are discontinuous and nonconvex, which needs to be gradually approximated with a sequence of semidefinite programming (SDP) problems. To approximate constraint $C_8$, we reformulate it with the rank constraints of two semidenifinite matrices. Specifically, we have $rank(\boldsymbol{P}) = |\mathcal{E}|$ if and only if there exists a $\boldsymbol{Z} \in \mathbb{S}^{n_b}$ such that
\begin{equation*}
    rank(\boldsymbol{Z}) = |\mathcal{E}|,
\end{equation*}
and
\begin{equation}
\label{eq:u_declare}
 rank\left(\boldsymbol{U} \right) \leqslant |\mathcal{E}|, \ 
 \boldsymbol{U} = \left[ \begin{matrix}
 \boldsymbol{I}_{|\mathcal{E}|} & \mathbf{P}^T \\
 \mathbf{P} & \boldsymbol{Z} \\
  \end{matrix} \right],
\end{equation}
where $\mathbb{S}^{n_b}$ is the set of symmetric $n_b \times n_b$ matrices. Based on the theorem in \cite{b22}, we have that when $e=0$ and $\boldsymbol{U}$ is a positive semidefinite matrix, $rank(\boldsymbol{Z}) = |\mathcal{E}|$ and $rank(\boldsymbol{U}) \leqslant |\mathcal{E}|$ are equivalent to

\begin{equation*}
V^{T}\boldsymbol{Z}V > 0,\ \text{and} \
    e\boldsymbol{I}_{n_b}-\boldsymbol{W}^T\boldsymbol{U}\boldsymbol{W}\succeq 0
\end{equation*}
where the vector 
$V\in \mathbb{R}^{n_b\times 1}$ is the eigenvector corresponding to the $n_b-|\mathcal{E}|+1$ smallest eigenvalue of $\boldsymbol{Z}$, and the matrices $\boldsymbol{W}\in \mathbb{R}^{(n_b+|\mathcal{E}|)\times n_b}$ are the eigenvectors corresponding to the $n_b$ smallest eigenvalues of $\boldsymbol{U}$.

Replacing the rank constraints with the SDP ones, we can solve $\mathcal{P}3$ by iteratively solving the following problem:
\begin{subequations}\label{eq:iterative_sdp_p3}
\begin{align}
(\mathcal{P}3\mbox{-}1) \min\limits_{\boldsymbol{P}^{(t)}, \boldsymbol{Z}^{t}, e^{t}} &\ \alpha\kappa(\mathbf{P})+\frac{(1-\alpha)}{P^*_{\mathrm{SOCP}}
}\sum_{i \in \mathcal{I}}\sum\limits_{(k,z) \in \mathcal{E}} p_k^{t}[i] + w^{t}e^{t} \label{eq:objective_p3}\\
\text{s.t.}\quad & \left(V^{t-1}\right)^{T}\boldsymbol{Z}^{t}V^{t-1} > 0, \label{eq:sdq_s} \\
& e^{t}\boldsymbol{I}_{n_b} - \left(\boldsymbol{W}^{t-1}\right)^T\boldsymbol{U}^{t}\boldsymbol{W}^{t-1}\succeq 0,  \label{eq:u_opt} \\
& 0 \leq e^{t} \leq e^{t-1}, \label{eq:sdp_e} \\
& C_5, C_6, C_7,  \\
& \boldsymbol{Z}^{t} \succeq \boldsymbol{0}, \boldsymbol{U}^{t} \succeq \boldsymbol{0},
\end{align}
\end{subequations}
where $p_k^{t}[i]$ is the transmit power of node $k$ at the $t$-th iteration,  $\boldsymbol{W}^{t-1}$ includes the eigenvectors corresponding to the $n_b$ smallest eigenvalues of $\boldsymbol{U}^{t-1}$, $w^{t}$ is a weighting factor increasing with the iteration count $t$, and $P^*_{\mathrm{SOCP}}$ is the power consumption of $\mathcal{P}2^*$ used to normalize the power consumption. The starting point at the $0$-th iteration can be obtained by solving Eq. (\ref{eq:iterative_sdp_p3}) without constraints (\ref{eq:sdq_s})-(\ref{eq:sdp_e}). At the 0-th iteration, $e^{0}$ is set to be the $n_b$-th smallest eigenvalue of $\boldsymbol{U}^{0}$.

To handle the nonconvexity of $\kappa(\mathbf{P})$, we express the condition number in terms of the singular values. The condition number of $\mathbf{P}$ is the ratio of its largest and smallest singular values, i.e., $k(\mathbf{P}) \triangleq \left(\frac{\lambda_{\max}(\mathbf{P}^T\mathbf{P})}{\lambda_{\min}(\mathbf{P}^T\mathbf{P})}\right)^{\frac{1}{2}}$.We can rewrite $\mathcal{P}3\mbox{-}1$ as an SDP problem, i.e.,
\begin{align}
(\mathcal{P}3\mbox{-}2) \ \min_{\mathbf{P},\gamma} \ & \alpha\gamma+\frac{(1-\alpha)}{P^*_{\mathrm{SOCP}}}\sum_{i \in \mathcal{I}}\sum\limits_{(k,z) \in \mathcal{E}} p_k^{t}[i] \nonumber \\
\text{s.t.} \ & C_5\mbox{-}C_8, C_{9}, \nonumber
\end{align}
where $C_{9}$ constrains the smallest and largest singular values of $\mathbf{P}$ as
\begin{align*}
 (C_{9})\  \mu \boldsymbol{I}_{|\mathcal{E}|}  \prec \mathbf{P}^T\mathbf{P} \prec \gamma \mu \boldsymbol{I}_{|\mathcal{E}|}.
\end{align*}
It is clear that this problem is nonlinear and nonconvex, which needs to be further simplified. We know from Eq. (\ref{eq:u_opt}) that when the iterative approach to $\mathcal{P}3$ converges, $e = 0$ and $rank(\boldsymbol{U}) \leq |\mathcal{E}|$. It has been shown in~\cite{b22} that $rank(\boldsymbol{U}) \leq |\mathcal{E}|$ indicates $rank(\boldsymbol{Z} - \mathbf{P}\mathbf{P}^T) = 0$, i.e., $\boldsymbol{Z} = \mathbf{P}\mathbf{P}^T$. Since it can be easily proved that the non-zero eigenvalues of $\mathbf{P}^T\mathbf{P}$ are also the eigenvalues of $\mathbf{P}\mathbf{P}^T$, we can replace $\mathbf{P}^T\mathbf{P}$ with $\mathbf{Z}$ in constraint $C_{10}$. Defining new variables $\nu=1/\mu$, $\tilde{Z}_{ij}=Z_{ij}/\mu$, $\tilde{P}_{ij}=P_{ij}/\mu$ and $\tilde{e}^t=e^t/\mu$,  we can rewrite $C_{10}$ as
\begin{equation*}
   (C_{10}^*) \ \begin{bmatrix}
 \boldsymbol{I}_{|\mathcal{E}|} & \mathbf{0} \\
\mathbf{0} &  \mathbf{0}
\end{bmatrix} \prec \mathbf{\tilde{Z}}^{t} \prec \gamma \begin{bmatrix}
 \boldsymbol{I}_{|\mathcal{E}|} & \mathbf{0} \\
\mathbf{0} &  \mathbf{0}
\end{bmatrix},
\end{equation*}
where $\mathbf{\tilde{Z}}^{t} = [\tilde{Z}_{ij}]$. Following the substitution, we can rewrite $C_5\textnormal{-}C_7$ as $C_5^*\textnormal{-}C_7^*$. Then, the problem $\mathcal{P}3\mbox{-}2$ can be solved by solving the following problem iteratively until convergence:
\begin{subequations}\label{eq:iterative_p3}
\begin{align}
\min\limits_{\substack{\boldsymbol{\tilde{P}}, \boldsymbol{\tilde{Z}}, \tilde{e}^{t}, \gamma, \nu}} &\alpha\gamma+\frac{(1-\alpha)}{P^*_{\mathrm{SOCP}}}\sum_{i \in \mathcal{I}}\sum\limits_{(k,z) \in \mathcal{E}} \tilde{p}_k^{t}[i]+ w^{t}\tilde{e}^{t} \\
\text{s.t.}\quad 
& C_5^*\textnormal{-}C_7^*, C_{9}^* \\
&\left(\tilde{V}^{t-1}\right)^{T}\boldsymbol{\tilde{Z}}^{t}\tilde{V}^{t-1} > 0, \\
& \tilde{e}^{t}\boldsymbol{I}_{n_b} - \left(\boldsymbol{\tilde{W}}^{t-1}\right)^T\boldsymbol{\tilde{U}}^{t}\boldsymbol{\tilde{W}}^{t-1}\succeq 0, \\
& 0 \leq \tilde{e}^{t} \leq \tilde{e}^{t-1}, \\
& \boldsymbol{\tilde{Z}}^{t} \succeq \boldsymbol{0}, \boldsymbol{\tilde{U}}^{t} \succeq \boldsymbol{0}, \nu > 0,
\end{align}
\end{subequations}
where $\boldsymbol{\tilde{U}} = \left[ \begin{matrix}
 \nu\boldsymbol{I}_{|\mathcal{E}|} & \mathbf{\tilde{P}}^T \\
 \mathbf{\tilde{P}} & \boldsymbol{\tilde{Z}} \\
  \end{matrix} \right], \ \mathbf{\tilde{P}} = [\tilde{P}_{ij}]$. $\tilde{V}^{t-1}$ and $\boldsymbol{\tilde{W}}^{t-1}$ are computed by $\boldsymbol{\tilde{Z}}^{t-1}$ and $\boldsymbol{\tilde{U}}^{t-1}$ as what $V^{t-1}$ and $\boldsymbol{W}^{t-1}$ to $\boldsymbol{Z}^{t-1}$ and $\boldsymbol{U}^{t-1}$, respectively. 

\subsection{Convergence Analysis}
\label{sec:convergence}
We first prove that the iterative approach for $\mathcal{P}2^*$ coverges. Let $(\boldsymbol{P}^t, \boldsymbol{\delta}^t, \boldsymbol{A}^t)$ be the solution of $\mathcal{P}2^*$ at the $t$-th iteration. At the $(t+1)$-th iteration, if we replace $(\boldsymbol{P}, \boldsymbol{\delta}, \boldsymbol{A})$ by $(\boldsymbol{P}^t, \boldsymbol{\delta}^t, \boldsymbol{A}^t)$, all the constraints in $\mathcal{P}2^*$ are still satisfied. This means that the solution at the $t$-th iteration is a feasible solution at the $(t+1)$-th iteration. As a result, the solution obtained at the $(t+1)$-th iteration is less or equal to that at the $t$-th objective. In addition that the solution of $\mathcal{P}2^*$ is bounded due to the power constraints, we have that $\mathcal{P}2^*$ is convergent. Since $\mathcal{P}3$ is solved iteratively, we want to prove the convergence of the iterative approaches. Specifically, we first want to prove that if there exists a solution at the $(i-1)$-th iteration, a solution will also exist at the $i$-th iteration, i.e., the problem is always solvable during the iterative process. Let $(\boldsymbol{U}^{t}, e^{t})$ be the solution to Eq. (\ref{eq:iterative_sdp_p3}) at the $t$-th iteration. Then, $(\boldsymbol{U}^{t}, e^{t})$ satisfies $e^{t}\boldsymbol{I}_{n_b} - \left(\boldsymbol{W}^{t-1}\right)^T\boldsymbol{U}^{t}\boldsymbol{W}^{t-1}\succeq 0 $, i.e., $e^{t} \geq \lambda_1 =  \lambda_{\max}\left(\left(\boldsymbol{W}^{t-1}\right)^T\boldsymbol{U}^{(t)}\boldsymbol{W}^{t-1}\right)$. We want to prove that $e^{t}\boldsymbol{I}_{n_b} - \left(\boldsymbol{W}^{t}\right)^T\boldsymbol{U}^{t}\boldsymbol{W}^{t}\succeq 0 $ is true, i.e., $e^{t} \geq \lambda_{\max}\left(\left(\boldsymbol{W}^{t}\right)^T\boldsymbol{U}^{t}\boldsymbol{W}^{t}\right)$, such that Eq. (\ref{eq:iterative_sdp_p3}) has at least a solution $(\boldsymbol{U}^{(t)}, e^{t})$ at the $(t+1)$-th iteration. By definition, we know the $n_b$-th smallest eigenvalue $\lambda_2 = \lambda_{\max}\left(\left(\boldsymbol{W}^{t}\right)^T\boldsymbol{U}^{t}\boldsymbol{W}^{t}\right)$. We next want to prove that $\lambda_1 \geq \lambda_2$. Recall that $\boldsymbol{W}^{t}$ includes the eignvectors corresponding to the $n_b$ smallest eigenvalues of $\boldsymbol{U}^{t}$. It has been proved in \cite{matrix_analysis} that for $\boldsymbol{W}^T\boldsymbol{W} = \mathbf{I}_{n_b}$, $\lambda_{\max}\left(\boldsymbol{W}^T\boldsymbol{U}^{t}\boldsymbol{W}\right) \geq \lambda_2$. Since $\left(\boldsymbol{W}^{t-1}\right)^T\boldsymbol{W}^{t-1} = \boldsymbol{I}_{n_b}$, we have $\lambda_1 \geq \lambda_2$. For the first iteration ($t = 1$), the pair $(\mathbf{U}^{0},e^{0})$ is a feasible solution for the iterative problem at the first iteration, where $e^0$ is the $(|\mathcal{E}|+1)$-th largest eigenvalue of $\mathbf{U}^{0}$. Since Eq. (\ref{eq:iterative_sdp_p3}) has a solution at each iteration and $e^{t} \leq e^{t-1}$, the iterative approach converges.

\section{Performance Evaluation}
\label{sec:perf_eval}

\begin{table}[tbp]
\centering  
\caption{Simulation Parameters}
\label{table:param}
\begin{tabular}{cc|cc} 
\hline
 \#Antenna per BS & 100 & BS distance & $\geq$ 15m \\
OFDM symbols per slot & 14 & Tx power range & 800-1200mW\\
Carrier frequency $f_c$ & 28GHz & Noise spectrum & -174dBm/Hz\\
Total bandwidth & 250MHz & BS density & $\geq$ 700BSs/km$^2$ \\
\hline
\end{tabular}
\vspace{-1em}
\end{table}

\subsection{Simulation Setup}

\begin{figure}[t]
    \centering
    \subfigure[Estimation error vs. weight]{
        \label{fig:weightchoice_error}
        \includegraphics[scale=0.35]{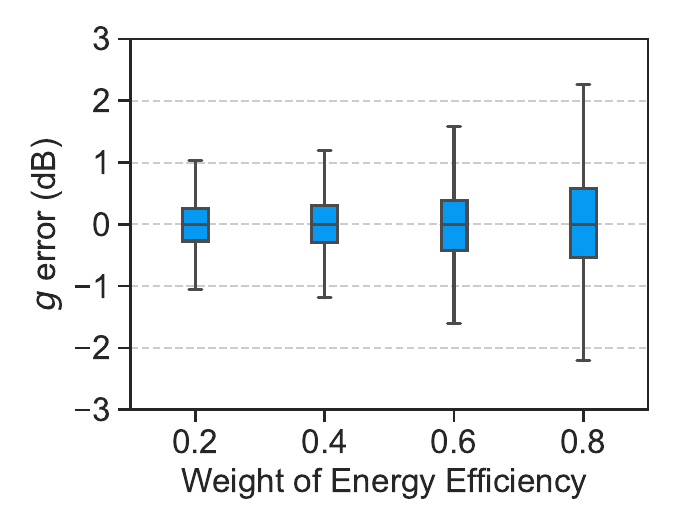}
    }
    \subfigure[Power consumption vs. weight]{
        \label{fig:weightchoice_overhead}
        \includegraphics[scale=0.35]{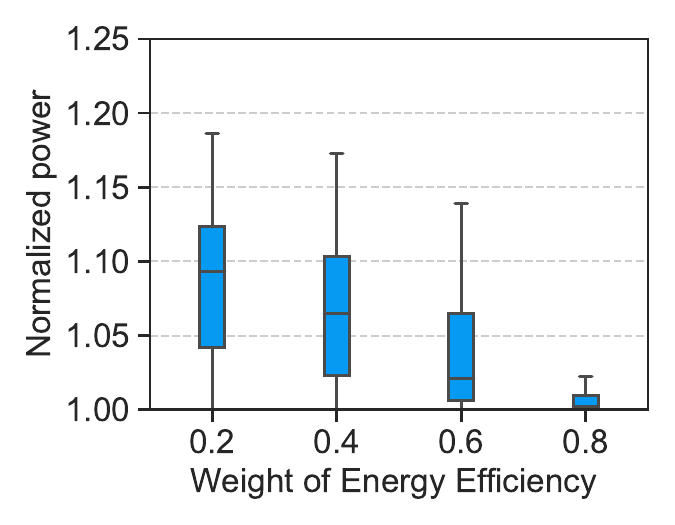}
    }
    \subfigure[Scheduling period length]{
        \label{fig:block_cn}
        \includegraphics[scale=0.34]{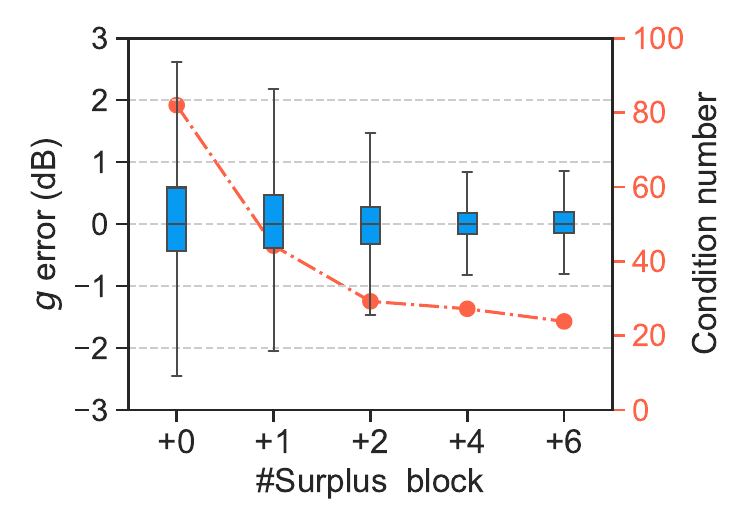}
    }
    \subfigure[Block length]{
        \label{fig:sample_g_error}
        \includegraphics[scale=0.34]{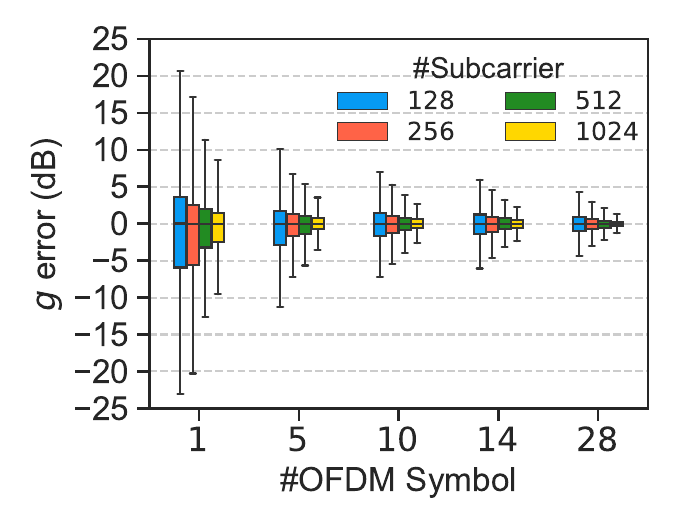}
    }
    \caption{Design parameter selection}
\end{figure}

The performance of our proposed joint optimization for IGE and resource allocation is evaluated via extensive simulations. A dense multi-hop mmWave backhaul network is simulated, where BSs are randomly located in a 0.01km$^2$ square field with density more than $700$ BSs/km$^2$. To prevent BSs from being either too close or distant, we control the distances between BSs within the range of 15m to 100m. To align with the 3GPP standards~\cite{TR38874}, we divide time into slots of equal length and each BS transmits data at the beginning of each time slot. Each time slot includes 14 OFDM symbols and the length of a time slot depends on the subcarrier spacing, ranging from 1ms down to 62.6$\mu$s.
BSs are synchronized with TOs and experience CFOs with the fractional component between $-0.5$ and $0.5$. Among all BSs, one is selected to serve as the gateway, while all other BSs backhaul traffic to/from the gateway. 
Between each BS and the gateway exists only a single path, determined by the Dijkstra algorithm based on the distances between BSs. The links between BSs are directional, allowing both upstream and downstream traffic, where the total number of links is twice the number of BSs minus one.
The mmWave channel $\mathbf{H}_{(k,z)}$ is modelled as a Rician channel, 
with a LOS component $\mathbf{H}_{(k,z)}^L$ and a scattering component $\mathbf{H}_{(k,z)}^S$, i.e., $\mathbf{H}_{(k,z)} =\sqrt{\frac{\varpi_{(k,z)}\zeta_{(k,z)} }{\zeta_{(k,z)} +1}}\mathbf{H}_{(k,z)}^L+\sqrt{\frac{\varpi_{(k,z)} }{\zeta_{(k,z)} +1}}\mathbf{H}_{(k,z)}^S$. The Rician K-factor $\zeta_{(k,z)}$ is set to 2 and $\varpi_{(k,z)}=10^{-(10\varrho  _1\mathrm{log}_{10}d_{(k,z)}+\varrho_2\mathrm{log}_{10}(4\pi/\lambda))/10}$, where $d_{(k,z)}$ is the distance of link $(k,z)$, $\lambda$ is the carrier wavelength, and $\varrho_1$, $\varrho_2$ are empirical propagation parameters as in \cite{zhao2018multi}. We simulate the randomness in the Rician channel by changing the scattering component $\mathbf{H}_{(k,z)}^S$, consisting of independent and identically distributed entries following Gaussian distribution as in~\cite{zhao2017multiuser,zhao2018multi}.

Each BS has access to a uniform linear array (ULA) with 100 antennas and the antenna spacing is equal to half of the wavelength. As a BS's transmitter is relatively close to its own receiver compared with the distance away from other nodes, we assume that SI channel suffers from Rayleigh flat fading without any effect of path loss, i.e. $(\mathbf{H}_z^{SI})_{i,j} \sim \mathcal{CN}(0, 1) $. Unless specified otherwise, we assume that the system has sufficient SI suppression with a SIC capability $\eta=-100$ dB according to \cite{ng2016power,li2021resource}. The angle-of-arrival (AoA) at each BS is the direction from which the strongest signal strength is detected, which are used to estimate the transmit and receive beamforming vectors of BSs. We allow each BS to control its transmit power between 800mW and 1200mW and use a different transmit power for each block. Each scheduling period is pre-determined by the number of BSs and the block length. We will discuss the choice of a proper block length in our experiments. The MCS for each link is determined based on the channel quality. The demand for each link takes a random number of time slots, less than the scheduling period, to send with the chosen modulation scheme. We use the square $M$-QAM modulation in our experiments and $M \in \{4, 16, 64, 256\}$. Each transmitter randomly selects modulated symbols from the constellation to send on each subcarrier. The simulation parameters are listed in Table \ref{table:param}. 


\subsection{Design Parameter Selection and Convergence}
\label{sec:design_param}

\begin{figure}
    \centering
    \includegraphics[scale=0.34]{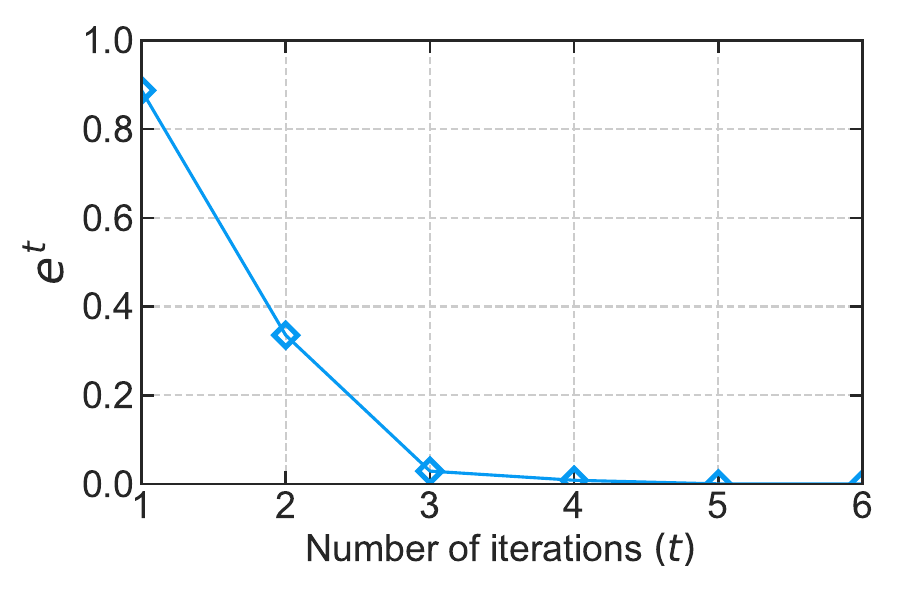}
    \caption{Convergence of our approach under a random 16-link topology}
    \label{fig:iteration}
\end{figure}

\begin{figure*}[t]
    \centering
    \subfigure[Estimation error vs. network size]{
        \label{fig:error_vs_net_size}
        \includegraphics[scale=0.41]{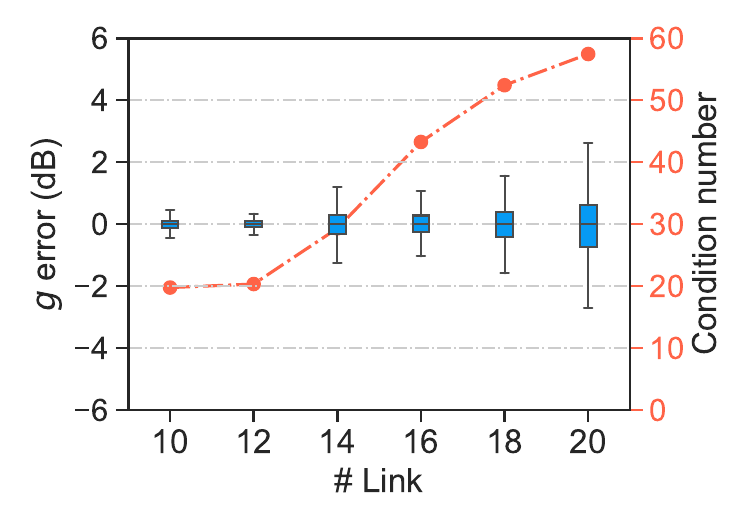}
    }
    \subfigure[Distribution of estimation errors]{
        \label{fig:h_size_estimation}
        \includegraphics[scale=0.4]{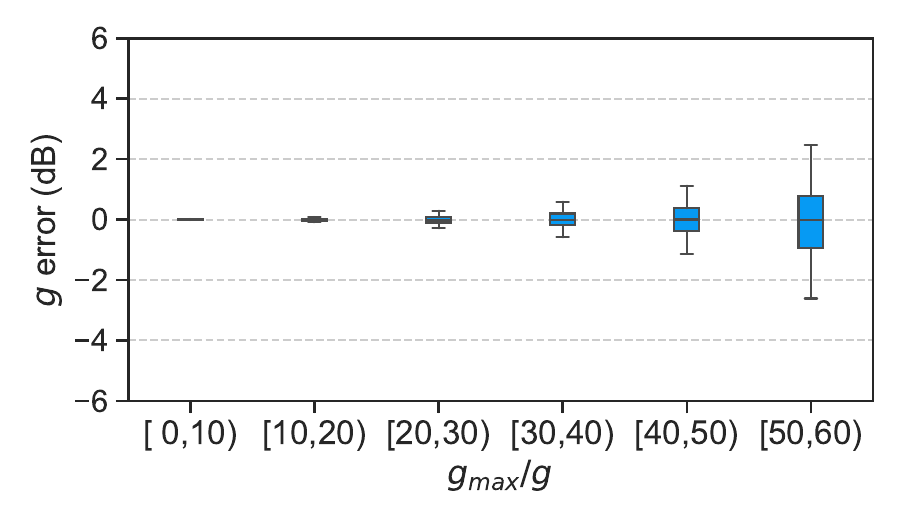}
    }
    \subfigure[Estimation errors by link type]{
        \label{fig:link_type_estimation}
        \includegraphics[scale=0.4]{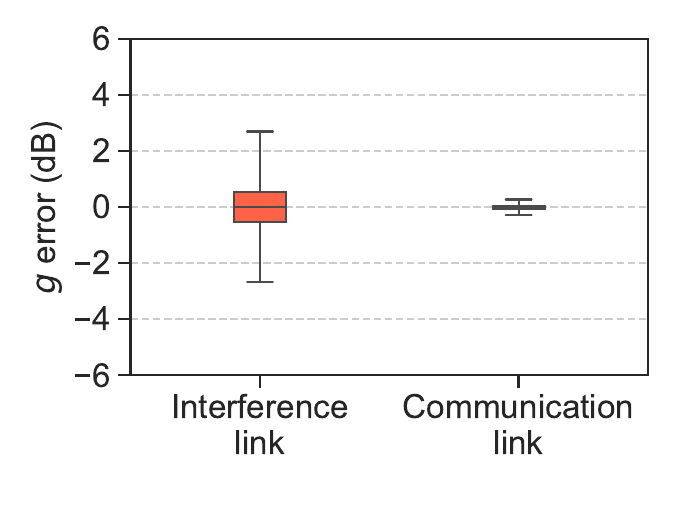}
    }
    \caption{The accuracy of equivalent channel gains is estimated with respect to network size, channel gain magnitude, and link type. In summary, both the interference and communication channel gains can be accurately estimated under different network sizes.}
    \vspace{-1em}
\end{figure*}

\noindent\textbf{Weight Choice for Multi-Objective Functions.}
To strike a balance between the channel gain estimation error and power consumption, we evaluate the two metrics under different weight choices. Since the experiments are conducted on random topologies, for better comparison, we normalize the total power consumption as in the objective function in Eq. (\ref{eq:objective_p3}). From Fig. \ref{fig:weightchoice_error} and \ref{fig:weightchoice_overhead}, we can see that as the weight of power consumption increases, the channel gain estimation error increases and the normalized power consumption decreases, as expected. Proper weights can be chosen based on the specific requirements on channel gain estimation and power consumption. In our experiments, we choose the power consumption weight of $0.8$ to control the extra power overhead attributed to IGE.

\noindent\textbf{Scheduling Period Length.}
As indicated by Theorem \ref{theo:cn_block}, a larger scheduling period helps reduce the condition number of the transmit power matrix. However, a larger scheduling period means a slower update of the channel estimation. We know that the minimum required number of blocks is equal to the number of BSs. The blocks exceeding the minimum required ones are referred to as \emph{surplus blocks}.
Fig. \ref{fig:block_cn} shows the estimation errors under different numbers of surplus blocks, where the network topologies are random with 8, 12, or 16 links. It can be seen that the median condition number decreases as the number of surplus blocks increases, which aligns with Theorem \ref{theo:cn_block}. As a result of the decreasing condition numbers, the estimation errors decrease sharply with several surplus blocks.
We thus use two surplus blocks in our experiments.

\noindent\textbf{Block Length.} 
As mentioned, we want to use the average transmit power to approximate the expected one with a sufficient number of samples. 
Fig. \ref{fig:sample_g_error} shows the distribution of the channel gain estimation errors under different numbers of subcarriers and OFDM symbols, where more than 100,000 channel gains are estimated under 16-link networks with random topologies for each configuration.
We can see that the estimation error decreases as more subcarriers and OFDM symbols are used, equating to using more samples in the time domain. When the number of subcarriers is 1024, the average absolute estimation error is as low as 0.6dB, even when a single OFDM symbol is used. There is no significant difference in estimation error when the number of OFDM symbols increases from 14 to 28. We thus set blocks to include 14 OFDM symbols, i.e., one time slot, in our experiments.


\noindent\textbf{Convergence.}
After all design parameters are determined, we evaluate the convergence performance of our approach under random topologies. As we discussed in Section \ref{sec:convergence}, when $e^t$ approaches to zero, the iterative approach to Eq. (\ref{eq:iterative_p3}) converges. Fig. \ref{fig:iteration} shows the convergence process under a random topology of 16 links. As the number of iterations $t$ increases, $e^t$ decreases to $0$. It can be seen that our iterative approach takes only 6 iterations to converge.







\subsection{Performance Evaluation}

\noindent\textbf{Channel Gain Estimation.} 
We measure the estimation error of the equivalent channel gain from link $(s,d)$ to $(k,z)$ as $10\log_{10}\left(\hat{g}_{(s,d),(k,z)}/g_{(s,d),(k,z)}\right)$ in dB, which is the log ratio of the estimated and actual equivalent channel gains. The ratio is equal to zero when the estimated and actual channel gains are equal and increases in its absolute value as the measured one deviates away from the actual one. Fig. \ref{fig:error_vs_net_size} shows the boxplot of the estimation errors using our approach under different network sizes, where the network size is measured by the number of links in the backhaul network. We can see that our approach achieves estimation errors less than 2dB for almost all the links, with half of the errors less than 0.8dB. 
The estimation errors slightly increase with the network size. As the network size increases, the condition number of the transmit power matrix tends to increase due to a larger dimension and results in larger estimation errors.

Fig. \ref{fig:h_size_estimation} shows the distribution of the channel gain estimation errors for 16-link networks with random topologies, where the relative magnitude of a link is calculated with respect to $g_{max}$, the maximum channel gain in the network. We can see that smaller channel gains tend to have larger estimation errors. For extremely small channel gains, the estimated values may fall below zero due to estimation errors. We choose to set such channel gains to the minimum channel gain in the network. Fortunately, such links with small equivalent channel gains are typically weak interference links, with signal strength much less than the communication links. Fig. \ref{fig:link_type_estimation} shows the estimation errors for communication and interference links, respectively. It can be seen that the interference links generally experience larger estimation errors than the communication links.
As a result, the estimated SINR is close to the actual one, ensuring that the joint optimization can perform as expected under estimation errors. 

\begin{figure}[t]
    \centering
    \subfigure[TO]{
        \label{fig:g_error_vs_to}
        \includegraphics[scale=0.36]{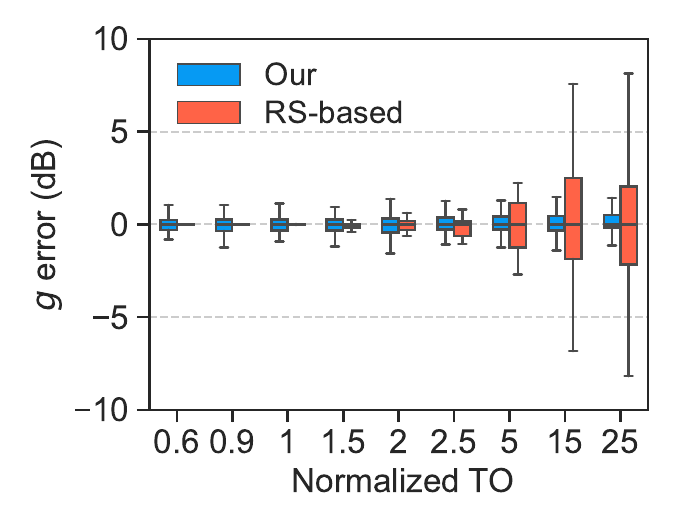}
    }
    \subfigure[CFO]{
        \label{fig:CDF_CFO}
        \includegraphics[scale=0.36]{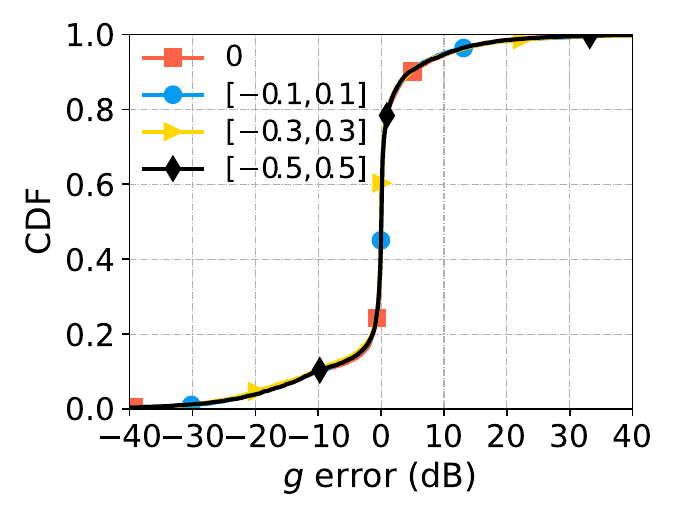}
    }
    \caption{Equivalent channel gains are estimated under TO and CFO. Our approach is robust to TO and CFO.}
\end{figure}

\noindent\textbf{Robustness to TO and CFO.}
Fig. \ref{fig:g_error_vs_to} shows the estimation errors under different TOs, where the TO is normalized with respect to CP length. 
Since our approach estimates channel gains in the power domain, it is robust to TO and achieves very small estimation errors under all TOs. Fig. \ref{fig:CDF_CFO} shows the estimation errors under different ranges of CFOs, where CFOs are randomly selected for each range. We can see that our approach achieves almost the same distribution of estimation errors as when no CFO exists (i.e., CFO = 0). This indicates that our approach is robust to CFOs. The channel gain estimation errors are mainly due to the randomness in received powers, not CFOs.



\begin{figure}
    \centering
    \includegraphics[scale=0.35]{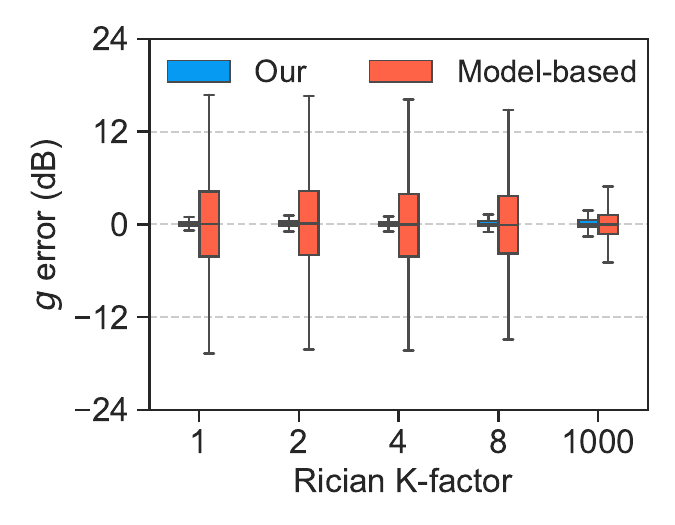}
    \caption{When the scattering component of the mmWave channel is large (small Rician K-factors), the model-based approach fails to accurately estimate the channel gains, while our approach performs well regardless of the K-factor.}
    \label{fig:model_based_approach}
\end{figure}

\textbf{Comparison with the Existing Methods.} We compare our approach with two existing methods for channel gain estimation: reference signal (RS)-based and model-based. The RS-based approach sends known reference symbols for measurement and the model-based approach is assumed using the same Rician channel model as in our experiments, where the K-factor is set to 2. We compare our approach and RS-based approach under different TOs. As shown in Fig. \ref{fig:g_error_vs_to}, the RS-based approach is accurate when TO is within CP and degrades rapidly when TO becomes greater than CP, while our approach is robust to TO. We compare our approach and the model-based approach under different Rician K-factors. Fig. \ref{fig:model_based_approach} shows that the model-based approach suffers from the randomness in the scattering component of the mmWave channel. Only when the LOS component is dominant (for large K-factors), the model-based approach can have small estimation errors as our approach does.

\begin{figure}
    \begin{minipage}{.24\textwidth}
        \includegraphics[scale=0.33]{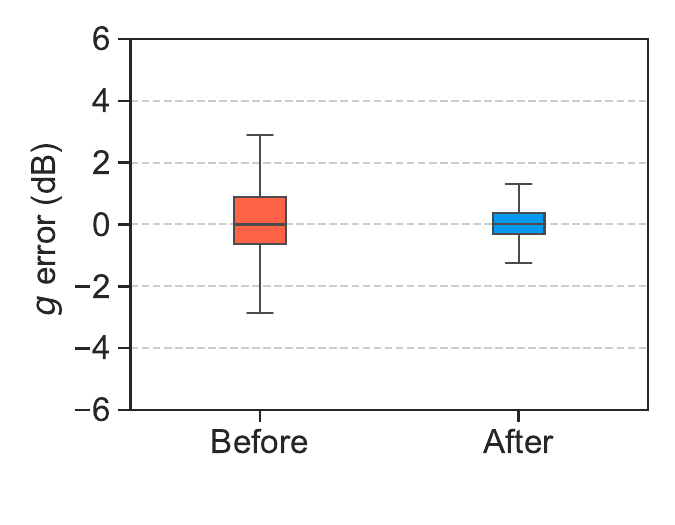}
        \caption{Impact of condition number}
        \label{fig:opt_cn}
    \end{minipage}%
    \hspace{0.12em}
    \begin{minipage}{0.24\textwidth}
            \includegraphics[scale=0.33]{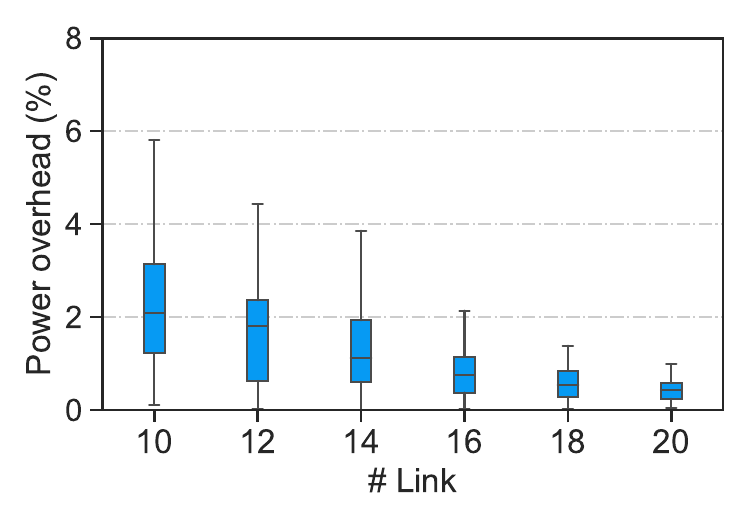}
        \caption{Power overhead}\label{fig:overhead_vs_net_size}
    \end{minipage}
\end{figure}

\noindent\textbf{Optimizing the Condition Number.} In our proposed solution, the second step is to minimize the condition number of the transmit power matrix. Fig. \ref{fig:opt_cn} shows the distributions of estimation errors before and after minimizing the condition number. We can see that reducing condition number helps reduce the estimation errors, consistent with what Lemma \ref{lemma:power_perturbation_one_channel} implies. Looking into how much condition numbers are reduced, we find that the reduction is significant for condition numbers of different magnitudes and that large condition numbers generally experience larger reductions than small ones.



\noindent\textbf{Power Overhead.} 
Our approach uses power as a new dimension for interference measurement and thus may have extra power overhead. We compare our approach with the pure resource allocation scheme not considering IGE (i.e., $\mathcal{P}2^*$) 
in terms of power consumption. Let $P_1$ and $P_0$ denote the power consumption in our approach and the pure power resource allocation scheme, respectively. We define \emph{power overhead} as $(P_1 - P_0) / P_0$. 
Fig. \ref{fig:overhead_vs_net_size} shows the boxplot for power overheads under different network sizes, where 100 random layouts are used for each network size. The positive power overheads indicate that our approach estimates the interference graph at the cost of power consumption. When the network includes 10 links, the 99-percentile power overhead is below 6\% and the average is about 2\%. The power overhead in general decreases as the network size increases because the extra power consumption needed for power control increases slower than the total power consumption of the network.

\begin{figure}
    \begin{minipage}{0.23\textwidth}
        \centering
        \includegraphics[scale=0.33]{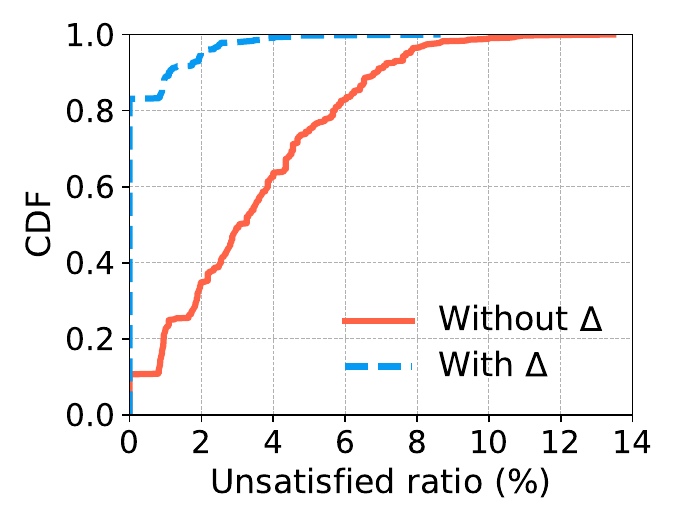}
        \caption{Robustness to channel gain estimation errors}
        \label{fig:SINR_Cdf}
    \end{minipage}
    \hspace{0.1em}
    \begin{minipage}{.23\textwidth}
        \centering
        \includegraphics[scale=0.33]{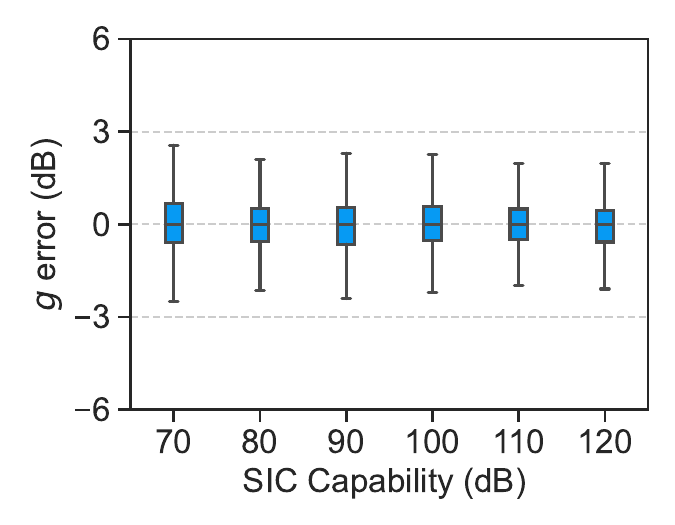}
        \caption{Channel gain estimation errors under different SIC capabilities}
        \label{fig:SIC_capability}
    \end{minipage}%
\end{figure}

\noindent\textbf{Robustness to Channel Estimation Errors.} 
To combat channel estimation errors, we have incorporated the error bounds of estimated channel gains in our problem formulation ($\mathcal{P}1$). We define the unsatisfied ratio as the percentage of links not satisfying the SINR constraints due to estimation errors. Fig. \ref{fig:SINR_Cdf} shows that the distributions of unsatisfied ratios with and without considering estimation errors ($\Delta$) under 700 random topologies. Compared with the power allocation without considering $\Delta$, our approach significantly reduces the unsatisfied ratios, with only 2\% of links slightly violating the SINR constraints for 5\% of random topologies. In contrast, about 60\% of topologies have an unsatisfied ratio greater than 2\% when $\Delta$ is not considered. The violation is because we use the  bounds of the expected channel estimation errors for power allocation. A more conservative bound can be used to further reduce the unsatisfied ratios.

\textbf{Impact of SIC Capability.} 
The above experiments are conducted under a fixed SIC capability of 100 dB. We re-do the above experiments for channel gain estimation errors under different SIC capabilities. Fig. \ref{fig:SIC_capability} shows that the distribution of the estimation errors does not have significant changes, even when the SIC capability is as low as 70 dB. This is because we consider the equivalent residual SI channels in the estimation and can accurately estimate their channel gains. Nonetheless, when a residual SI channel is overly strong and significantly affects the overall magnitude of the received signal, it may affect the estimation accuracy of weak channels as indicated by the error analysis in Section \ref{sec:error_analysis}.

\begin{table}[tbp]
\centering  
\caption{Comparison with the B\&B approach}
\label{table:BnB}
\begin{tabular}{l|c|c|c|c} 
\hline
 & 4 links & 8 links & 12 links & 16 links \\
 \hline
Power Ratio & 1.03 & 1.03 & 1.02 & 1.02 \\
Run Time Ratio & 0.07 & 0.05 & 0.02 & 0.006 \\
\hline
\end{tabular}
\vspace{-1em}
\end{table}

\textbf{Optimality Loss.} We compare our approach with the B\&B approach that aims to find the optimal solution for resource allocation in terms of power consumption and run time. Table \ref{table:BnB} shows the experimental results under four different network sizes, where the power and run time ratios are the power consumption and run time of our approach divided by those of the B\&B approach, respectively. It can be seen that the power ratio is close to one, indicating that our approach experiences no significant optimality loss for power consumption. Meanwhile, our approach takes only a few percent of the run time of the B\&B approach and the gap in run time continues to grow as the network size increases.

\section{Conclusion}
\label{sec:conclusion}

In this paper, we proposed to use power as a new dimension for interference graph estimation, such that it can be done simultaneously with resource allocation using the same frequency-time resources. To this end, we proved that the mmWave backhaul network can be considered as a linear system such that the interference graph can be measured with power control. We derived bounds for the estimation errors of channel gains and found that minimizing condition number of the transmit power matrix is effective in reducing the estimation errors. We formed a multi-objective joint optimization of the resource allocation and IGE, and showed that our approach is robust to timing and carrier frequency offsets. Moreover, the overhead of our approach was shown to be minimal by experiments. As for future work, we will
apply the joint optimization framework to other wireless networks.





\appendices
\section{Proof of Lemma \ref{lemma:rx_power_linearity}}
\label{section:rx_power_linearity}
Since $\bm{v}[i] \sim \mathcal{CN}(\bm{0}, \sigma^2_v\bm{\textnormal{I}})$, $\mathbb{E}[n_z[i]] = \mathbb{E}[\bm{\widetilde{\omega}}^T_{z}\bm{v}[i]] = 0$ and $\mathbb{E}\left[|n_z[i]|^2\right] = \bm{\widetilde{\omega}}^T_{z}\mathbb{E}\left[\bm{v}[i]\bm{v}[i]^H\right]\bm{\widetilde{\omega}}^*_{z} =\sigma^2_v$. 
Let $F_{(s, d),(k,z)}[i] =  h^{\textnormal{eq}}_{(s,d),(k,z)} x_s[i - \mu_{(s,d)}]$. Based on Eq. (\ref{eq:simplified_received_signal}), the expected receive power of node $z$ can be calculated by omitting the items multiplying $n_z[i]$ as
\begin{align*}
    &\mathbb{E}[|y_z[i]|^2]  \\
    &= \sum_{(s,d)\in\mathcal{E}} |h^{\textnormal{eq}}_{(s,d),(k,z)}|^2\mathbb{E}\left[\left|x_s[i-\mu_{(s,d)}]\right|^2\right] \\
    &\quad
    + \sum_{(s,d)\in\mathcal{E}}\sum_{\substack{(l,v)\in\mathcal{E} \\ (l,v)\not=(s,d)}} \mathbb{E}\left[F_{(s,d), (k,z)}[i]F^*_{(l,v),(k,z)}[i]\right] + \sigma_v^2.
\end{align*}
Since $x_s[i]$ and $x_l[i]$ are independent for $s\not=l$, we have $\mathbb{E}[x_s[i]x_l[i]^*] = \mathbb{E}[x_s[i]]\mathbb{E}[x_l[i]^*] = 0$, where $\mathbb{E}[x_s[i]] = 0$ for all $s$'s, because $x_s[i] = \frac{1}{\sqrt{N_c}}\sum_{r=0}^{N_c-1}X_s[r]e^{j2\pi ir/N_c}$ and $\mathbb{E}[X_s[r]] = 0$ for $1 \leq m \leq N_c$. This implies that when $(l,v)\not=(s,d)$,
\begin{align*}
    &\mathbb{E}\left[F_{(s,d),(k,z)}[i]F^*_{(l,v),(k,z)}[i]\right] \\
    & = q\mathbb{E}\left[x_s[i-\mu_{(s,d)}]\right]\mathbb{E}\left[x_l^*[i-\mu_{(l,v)}]\right] = 0,
\end{align*}
where $q = h^{\textnormal{eq}}_{(s,d),(k,z)}\left(h^{\textnormal{eq}}_{(l,v),{(k,z)}}\right)^*\mathbb{E}_{\Delta\phi}\left[e^{j2\pi i\Delta\phi}\right]$ and $\Delta\phi = \phi_{(s,z)}-\phi_{(l,z)}$. 

\section{Proof of Theorem \ref{theorem:bound}}
\label{section:proof_theorem_bound}

Since the I/Q components of $x_k[i]$ are assumed to be independent and that the modulated symbols for each component are selected from the constellation equi-probably, we have that $\mathbb{E}[X^I_k[i]] = \mathbb{E}[X^Q_k[i]] = 0$ and $\sigma^2_k = \Var(X^I_k[i]) = \Var(X^Q_k[i])$. Suppose we want to calculate the average power of received signals over $N_k$ OFDM symbols. As each OFDM symbol includes $N_s = N_c + N_g$ time-domain samples, the average signal power is $\bar{p}^{tx}_k = \frac{1}{N_kN_s}\sum_{j=1}^{N_kN_s}|x_k[j]|^2$. We want to apply the concentration inequalities to bound the difference between $\bar{p}^{tx}_k$ and $\mathbb{E}[|x_k[i]|^2]$. However, the concentration inequalities generally assume the independence between variables, while the transmitted signals, $x_k[i]$'s, are the IDFTs of all modulated symbols on $N_c$ subcarriers and are thus dependent. We thus want to relate $x_k[i]$ with the modulated symbols, $X_k[z]$'s, that are independently chosen.

Based on the Parseval's Theorem for discrete signals, the sum of transmitted signals within an OFDM symbol is
\begin{equation}\label{eq:parseval}
    \sum_{i=0}^{N_s-1} |x_k[i]|^2 = \sum_{z=0}^{N_c-1}|X_k[z]|^2 + \sum_{i=0}^{N_g-1}|x_k[i]|^2.
\end{equation}
As $x_k[i]$ is the IDFT of $X_k[z]$'s, it can be written as
\begin{align*}
    |x_k[i]|^2 &= \frac{1}{N_c}\sum_{d=0}^{N_c-1}\sum_{z=0}^{N_c-1}X_k[d]X_k[z]^*e^{j2\pi(d-z)i/N}, \\
    &= \frac{1}{N_c}\sum_{d=0}^{N_c-1}|X_k[d]|^2 + \frac{2}{N_c}\sum_{d=1}^{N_c-1}\sum_{z=0}^{d-1} Re\{X_k[d]X_k[z]^*e^{j\theta}\},
\end{align*}
where $\theta = 2\pi(d-z)i/N$.
Then, this implies that 
\begin{equation*}
\mathbb{E}[|x_k[i]|^2] = \mathbb{E}[|X_k[z]|^2] = \mathbb{E}[(X^I_k[z])^2] + \mathbb{E}[(X^Q_k[z])^2] = 2\sigma^2_k,    
\end{equation*} 
and Eq. (\ref{eq:parseval}) can be rewritten as
\begin{align*}
\sum_{i=0}^{N_s-1} |x_k[i]|^2 = \frac{N_{s}}{
    N_c}\sum_{z=0}^{N_c-1}|X_k[z]|^2 + \frac{2}{N_c}\sum_{i=0}^{N_g-1}\sum_{d=1}^{N_c-1}\sum_{z=0}^{d-1} f_k(X),
\end{align*}
where $f_k(X) = Re\{X_k[d]X_k[z]^*e^{j\theta}\}$.

As $X_k[d] = X^I_k[d] + jX^Q_k[d]$, we can rewrite $f_k(X)$ as
\begin{align*}
    f(X_k) = &X^I_k[d]X^I_k[z]\cos(\theta) + X_k[d]^QX_k^Q[z]\cos(\theta) \\
               &- X^Q_k[d]X^I_k[z]\sin(\theta) + X^I_k[d]X^Q_k[z]\sin(\theta),
\end{align*}
where $d \not= z$. For zero-mean independent variables $x_i$'s, we have 
\begin{align*}
    \Var(x_1\dots x_d) &= \prod_{i=1}^d\left(\Var(x_i) + \mathbb{E}^2[x_i]\right) - \prod_{i=1}^{d}\mathbb{E}^2[x_i] \\
    &= \prod_{i=1}^d\Var(x_d),
\end{align*}
which implies $\Var(X_k^I[z]X_k^I[d]) = \Var(X_k^I[z])\Var(X_k^I[d])$. In addition that $\Var(X^I_k[z]\cos(\theta)) = \mathbb{E}[(X^I_k[z])^2]\cos^2(\theta)$ and $\Var(X^I_k[d]) = \Var(X^Q_k[d])$, we can have $\Var(f(X_k))=2\sigma_k^4$. To use the Bennett's inequality, we further need to bound $f_k(X)$ as
\begin{equation*}
    \left|f_k(X)\right| \leq \left|X_k[d]\right|\left|X_k[z]\right|\left|e^{j\theta}\right| \leq B_{k,1} = 2\left(X^{I,max}_k\right)^2.
\end{equation*}
Combining the above facts, $f(X_k)$ can be bounded as
\begin{equation*}
    \mathbb{P}\left[\frac{1}{n}\sum_{i=0}^{n-1}f(X_k) \geq \Delta\right] \leq \exp\left(-\frac{n\sigma_f^2}{B_{k,1}^2}h\left(\frac{B_{k,1}\Delta}{\sigma_f^2}\right)\right),
\end{equation*}
where $n$ is the number of samples and $\sigma_f^2 = \Var(f(X_k))$.

Let $\mu_k = \mathbb{E}[|x_k[i]|^2]$ and $0 \leq \delta \leq 1$ denote the approximation error between $\bar{p}^{tx}_k$ and $\mu_k$. The probability of $\bar{p}^{tx}_k$ being greater than $\mu_k(1+\delta)$ is
\begin{align*}
    &\mathbb{P}\left[\bar{p}^{tx}_k - \mu_k \geq \mu_k\delta \right] \\ 
    &
    = \mathbb{P}\left[\frac{1}{N_kN_c}\sum_{i=0}^{N_k-1}\sum_{z=0}^{N_c-1}|X_k[z+N_ci]|^2 - \mu_k \geq \mu_k\delta - f(X) \right] \\
    &
    \leq \mathbb{P}\left[f(X) \leq \Delta \right]\mathbb{P}[E] + \mathbb{P}[f(X) \geq \Delta],
\end{align*}
where 
\begin{equation*}
   f(X) = \frac{2}{N_kN_sN_c}\sum_{i=0}^{N_k-1}\sum_{l=0}^{N_g-1}\sum_{d=1}^{N_c-1}\sum_{z=0}^{d-1}f_k(X),
\end{equation*}
and 
\begin{equation*}
   \mathbb{P}[E] = \mathbb{P}\left[\frac{1}{N_kN_c}\sum_{i=0}^{N_k-1}\sum_{z=0}^{N_c-1}|X_k[z+N_ci]|^2 - \mu_k \geq \mu_k\delta - \Delta \right].
\end{equation*}
Based on the Bennett's inequality, we have
\begin{equation*}
    \mathbb{P}[E] \leq \exp\left(-\frac{N_kN_c\sigma_S^2}{B_{k,2}^2}h\left(\frac{B_{k,2}(\mu_k\delta - \Delta)}{\eta_k^2}\right)\right).
\end{equation*}
where $B_k$ and $\eta_k^2$ are the upper bound and variance for $|X_k[z]|^2 - \mu_k$, respectively. Since $\mu_k = 2\sigma_k^2$ and $X^{I,max}_k \geq 2\sigma_k^2$, $B_k = 2\left(X^{I,max}_k\right)^2 - 2\sigma_k^2$, and $\eta_k^2 = \mathbb{E}[|X_k[z]|^4] - 4\sigma_k^4$. Moreover, we know that $\bar{p}^{tx}_k - \mu_k \leq -\mu_k\delta$ if and only if $-\bar{p}^{tx}_k + \mu_k \geq \mu_k\delta$, which implies that $\mathbb{P}\left[\left|\bar{p}^{tx}_k - \mu_k\right| \geq \mu_k\delta \right] \leq 2\mathbb{P}\left[\bar{p}^{tx}_k - \mu_k \geq \mu_k\delta \right]$.

\section{Proof of Theorem \ref{eq:f_upper_bound}}
\label{appendix:f_upper_bound}

We want to compute a bound for each item using the Chebyshev's inequality, which requires knowing the variance of the variable. It has been shown in \cite{ochiai2001distribution} that $x_s[i]$ approximately follows the complex Gaussian distribution when the number of subcarriers is large based on the central limit theorem. Since $x_s[i-\mu_{(s,d)}]$ have same distribution with $x_s[i]$ and recall Eq.(\ref{eq:Modulation_with_time_series}),  \cite{ochiai2001distribution} has shown it become complex gaussian with variance $2\sigma^2_s$ ($\sigma^2_s$ is the variance of modulated symbol $X_s$'s real/imaginary part) by the central limit theorem. As a result, $\mathbb{E}[|x_s[r]|^2]=2\sigma_{s}^2$.

Let $D_{(k,z)}[r]=F_{(k,z)}[r]+V_{(k,z)}[r]$, which is a real-valued variable. We first calculate $\mathbb{E}[D^2_{(k,z)}[r]]$ for item $\circled{\small \textnormal{1}}$. As $\mathbb{E}_x[x_s[r]] = 0$, $\mathbb{E}_x[q_{(s,d),(k,z)}[r]] = 0$ and $\mathbb{E}_x[q^2_{(s,d),(k,z)}[r]]=0$. We thus have that 
\begin{align*}
    \label{eq:D_expected}
    \begin{split}    
    Var_{x}(\circled{\small\textnormal{1}})&=\mathbb{E}_{x}[\circled{\small 1}^2]-(\mathbb{E}_{x}[\circled{\small 1}])^2\\
    &=\frac{1}{m_1^2}\mathbb{E}_x[(\sum_{r=(m_1-1)i+1}^{m_{1}i}D_{(k,z)}[r])^{2}]\\ &=\frac{1}{m_1}\sum_{e_1\in\mathcal{E}}\sum_{\substack{e_2\in\mathcal{E} \\ e_2\not=e_1}}\mathbb{E}_x[|q_{e_1, (k,z)}[r]|^2]\mathbb{E}_x[|q^*_{e_2,(k,z)}[r]|^{2}] \\
    &=\frac{4}{m_1}\sum_{e_1\in\mathcal{E}}\sum_{\substack{e_2\in\mathcal{E} \\ e_2\not=e_1}} |h^{\textnormal{eq}}_{e_1,(k,z)}|^2|h^{\textnormal{eq}}_{e_2,(k,z)}|^{2} \sigma^2_s\sigma^2_l.     
    \end{split}
\end{align*}
where $e_1=(s,d)$ and $e_2=(l,v)$.
Using the Chebyshev's inequality, we have $\mathbb{P}[|\circled{\small \textnormal{1}}| > A_1] \leq \frac{Var(\circled{\small \textnormal{1}})}{A_1^2}$.

For item $\circled{\small 2}$, we know $\mathbb{E}[\circled{\small 2}]=0$ and 
\begin{align*}
    Var_{x}(\circled{\small\textnormal{2}})&=\mathbb{E}_{x}[\circled{\small 2}^2]-(\mathbb{E}_{x}[\circled{\small 2}])^2\\
    &=-4(\sum_{(s,d)\in \mathcal{E}}|h_{(s,d),(k,z)}^{eq}|^2\sigma_s^2)^2+\\
    &\frac{1}{m_2^2}\mathbb{E}[(\sum_{r=(m_2-1)i+1}^{m_2i}\sum_{(s,d)\in\mathcal{E}}|h_{(s,d),(k,z)}^{eq}|^2|x_s[r]|^2)^2]\\
    &=\frac{1}{m_2}\sum_{(s,d)\in\mathcal{E}}|h_{(s,d),(k,z)}^{eq}|^4(\mathbb{E}_x[|x_s[r]|^4]-4\sigma_s^4)
\end{align*}

Using Eq.  (\ref{eq:Modulation_with_time_series}), we can compute 
\begin{align*}
    \mathbb{E}_x[|x_s[r]|^4]&=\frac{1}{N_c^2}\sum_{i=0}^{N_c-1}\sum_{j=0}^{N_c-1}|X_s[i]|^2|X_s[j]|^2+\\
    &\frac{1}{N_c^2}\sum_{i=0}^{N_c-1}\sum_{\substack{j=0\\j\ne i}}^{N_c-1}|X_s[i]|^2|X_s[j]|^2\\
    &=8\sigma_s^4-\frac{4}{N_c}\sigma_s^4
\end{align*}
Using the Chebyshev's inequality, we have
$\mathbb{P}[|\circled{\small \textnormal{2}}| > A_2] \leq \frac{Var(\circled{\small \textnormal{2}})}{A_2^2}$

For item $\circled{\small 3}$, since $\bm{v}[i] \sim \mathcal{CN}(\bm{0}, \sigma^2_v\bm{\textnormal{I}})$, and $n_z[i] = \widetilde{\bm{\omega}}^T_{(k,z)}\bm{v}[i]$, we have $n_z[i] \sim \mathcal{CN}(0, \sigma^2_v)$. Thus, $\mathbb{E}_{\widetilde{v}}[\circled{\small 3}]=\sigma^2_v$ and
\begin{align*}
     Var(\circled{\small 3})&=\mathbb{E}_{n}[\circled{\small 3}^2]-(\mathbb{E}_{n}[\circled{\small 3}])^2\\
   &=\frac{1}{m_3^2}\mathbb{E}_{n}[(\sum_{r=(m_3-1)i+1}^{m_3i}(n_z^{I}[r])^2+(n_z^Q[r])^2)^2]-\sigma_{v}^4\\
   &=\frac{1}{m_3}\sigma_v^4.   
\end{align*}
Using the Chebyshev's inequality, we have $\mathbb{P}[|\circled{\small \textnormal{3}}-\sigma_v^2| > A_3] \leq \frac{Var(\circled{\small \textnormal{3}})}{A_3^2}=\frac{\sigma_v^4}{m_3A_3^2}$.

\bibliographystyle{IEEEtran}
\bibliography{IEEEabrv}

\end{document}